\documentclass[sigconf]{acmart}
\usepackage[norelsize,ruled,vlined,linesnumbered]{algorithm2e}
\usepackage{algorithmic}
\usepackage{subfigure}

\AtBeginDocument{%
  \providecommand\BibTeX{{%
    \normalfont B\kern-0.5em{\scshape i\kern-0.25em b}\kern-0.8em\TeX}}}

\copyrightyear{2021}
\acmYear{2021}
\setcopyright{acmcopyright}
\acmConference[]{}{}
\acmBooktitle{}
\acmPrice{15.00}
\acmDOI{10.1145/3460231.3474229}
\acmISBN{978-1-4503-8458-2/21/09}

\setcopyright{none}
\renewcommand\footnotetextcopyrightpermission[1]{} 
\newcommand{\wsq}{\hspace{\fill}$\square$}
\newcommand{\argmax}{\mathop{\rm arg\,max}\limits}

\newtheoremstyle{mystyle}
    {1.5mm}
    {1.5mm}
    {\it}
    {0mm}
    {\scshape}
    {.}
    { }
    {}
\theoremstyle{mystyle}
\newtheorem{defi}{Definition}
\newtheorem{exa}{Example}
\newtheorem{lemm}{Lemma}
\newtheorem{theo}{Theorem}
\newtheorem{coro}{Corollary}

\begin{document}

\title{Reverse Maximum Inner Product Search: How to efficiently find users who would like to buy my item?}

\author{Daichi Amagata}
\affiliation{%
  \institution{Osaka University, JST PRESTO}
  \country{Japan}}
\email{amagata.daichi@ist.osaka-u.ac.jp}

\author{Takahiro Hara}
\affiliation{%
  \institution{Osaka University}
  \country{Japan}
}
\email{hara@ist.osaka-u.ac.jp}

\renewcommand{\shortauthors}{Daichi Amagata and Takahiro Hara}

\begin{abstract}
The MIPS (maximum inner product search), which finds the item with the highest inner product with a given query user, is an essential problem in the recommendation field.
It is usual that e-commerce companies face situations where they want to promote and sell new or discounted items.
In these situations, we have to consider a question: who are interested in the items and how to find them?
This paper answers this question by addressing a new problem called reverse maximum inner product search (reverse MIPS).
Given a query vector and two sets of vectors (user vectors and item vectors), the problem of reverse MIPS finds a set of user vectors whose inner product with the query vector is the maximum among the query and item vectors.
Although the importance of this problem is clear, its straightforward implementation incurs a computationally expensive cost.

We therefore propose Simpfer, a simple, fast, and exact algorithm for reverse MIPS.
In an offline phase, Simpfer builds a simple index that maintains a lower-bound of the maximum inner product.
By exploiting this index, Simpfer judges whether the query vector can have the maximum inner product or not, for a given user vector, in a constant time.
Besides, our index enables filtering user vectors, which cannot have the maximum inner product with the query vector, in a batch.
We theoretically demonstrate that Simpfer outperforms baselines employing state-of-the-art MIPS techniques.
Furthermore, our extensive experiments on real datasets show that Simpfer is at least two orders magnitude faster than the baselines.
\end{abstract}

\maketitle

\section{Introduction}  \label{section_introduction}
The MIPS (maximum inner product search) problem, or $k$-MIPS problem, is an essential tool in the recommendation field.
Given a query (user) vector, this problem finds the $k$ item vectors with the highest inner product with the query vector among a set of item vectors.
The search result, i.e., $k$ item vectors, can be used as recommendation for the user, and the user and item vectors are obtained via Matrix Factorization, which is well employed in recommender systems \cite{cremonesi2010performance, van2013deep, chen2020efficient, fraccaro2016indexable}.
Although some learned similarities via MLP (i.e., neural networks) have also been devised, e.g., in \cite{zamani2020learning,zhao2020improving}, \cite{rendle2020neural} has actually demonstrated that inner product-based (i.e., Matrix Factorization-based) recommendations show better performances than learned similarities.
We hence focus on inner product between $d$-dimensional vectors that are obtained via Matrix Factorization.

\subsection{Motivation}
The $k$-MIPS problem is effective for the case where a user wants to know items that s/he prefers (i.e., user-driven cases), but e-commerce companies usually face situations where they want to advertise an item, which may be new or discounted one, to users, which corresponds to item-driven cases.
Trivially, an effective advertisement is to recommend such an item to users who would be interested in this item.

In the context of the $k$-MIPS problem, if this item is included in the top-k item set for a user, we should make an advertisement of the item to this user.
That is, \textit{we should find a set of such users}.
This paper addresses this new problem, called \textit{reverse $k$-MIPS problem}.
To ease of presentation, this section assumes that $k = 1$ (the general case is defined in Section \ref{section_preliminary}).
Given a query vector $\mathbf{q}$ (the vector of a target item) and two sets of $d$-dimensional vectors $\mathbf{Q}$ (set of user vectors) and $\mathbf{P}$ (set of item vectors), the reverse MIPS problem finds all user vectors $\mathbf{u} \in \mathbf{Q}$ such that $\mathbf{q} =$ arg\,max$_{\mathbf{p} \in \mathbf{P} \,\cup\, \{\mathbf{q}\}}\mathbf{p} \cdot \mathbf{u}$.

\begin{exa} \label{example_rmips}
Table \ref{table_example} illustrates $\mathbf{Q}$, $\mathbf{P}$, and the MIPS result, i.e., $\mathbf{p^{*}} =$ {\rm arg\,max}$_{\mathbf{p} \in \mathbf{P}}\,\mathbf{u} \cdot \mathbf{p}$, of each vector in $\mathbf{Q}$.
Let $\mathbf{q} = \mathbf{p_{5}}$, and the result of reverse MIPS is $\{\mathbf{u_{3}},\mathbf{u_{4}}\}$ because $\mathbf{p_{5}}$ is the top-1 item for $\mathbf{u_{3}}$ and $\mathbf{u_{4}}$.
When $\mathbf{q} = \mathbf{p_{1}}$, we have no result, because $\mathbf{p_{1}}$ is not the top-1 item $\forall \mathbf{u} \in \mathbf{Q}$.
Similarly, when $\mathbf{q} = \mathbf{p_{2}}$, the result is $\{\mathbf{u_{2}}\}$.
\end{exa}

\noindent
From this example, we see that, if an e-commerce service wants to promote the item corresponding to $\mathbf{p_{5}}$, this service can obtain the users who would prefer this item through the reverse MIPS, and sends them a notification about this item.

The reverse $k$-MIPS problem is an effective tool not only for item-driven recommendations but also market analysis.
Assume that we are given a vector of a new item, $\mathbf{q}$.
It is necessary to design an effective sales strategy to gain a profit.
Understanding the features of users that may prefer the item is important for the strategy.
Solving the reverse $k$-MIPS of the query vector $\mathbf{q}$ supports this understanding.

\begin{table}[!t]
    \centering
	\caption{Example of reverse MIPS where $\mathbf{q} = \mathbf{p_{5}}$. The rows at right illustrate the result of MIPS on $\mathbf{P}$ for each $\mathbf{u} \in \mathbf{Q}$.}	\label{table_example}
	\begin{tabular}{c|c||c|c||c|c} \hline
        \multicolumn{2}{c||}{$\mathbf{Q}$}              & \multicolumn{2}{c||}{$\mathbf{P}$}             & \multicolumn{2}{c}{$\mathbf{p^{*}}$ (MIPS result)}   \\ \hline \hline
        $\mathbf{u_{1}}$    & $\langle 3.1, 0.1\rangle$ & $\mathbf{p_{1}}$  & $\langle 2.8, 0.6\rangle$  & $\mathbf{u_{1}}$ & $\mathbf{p_{3}}$                  \\ \hline
        $\mathbf{u_{2}}$    & $\langle 2.5, 2.0\rangle$ & $\mathbf{p_{2}}$  & $\langle 2.5, 1.8\rangle$  & $\mathbf{u_{2}}$ & $\mathbf{p_{2}}$                  \\ \hline
        $\mathbf{u_{3}}$    & $\langle 1.5, 2.2\rangle$ & $\mathbf{p_{3}}$  & $\langle 3.2, 1.0\rangle$  & $\mathbf{u_{3}}$ & $\mathbf{p_{5}}$                  \\ \hline
        $\mathbf{u_{4}}$    & $\langle 1.8, 3.2\rangle$ & $\mathbf{p_{4}}$  & $\langle 1.4, 2.6\rangle$  & $\mathbf{u_{4}}$ & $\mathbf{p_{5}}$                  \\ \hline
                            &                           & $\mathbf{p_{5}}$  & $\langle 0.5, 3.4\rangle$  &                  &                                   \\ \hline
	\end{tabular}
\end{table}

\subsection{Challenge}
The above practical situations clarify the importance of reverse MIPS.
Because e-commerce services have large number of users and items, $|\mathbf{Q}|$ and $|\mathbf{P}|$ are large.
In addition, a query vector is not pre-known and is specified on-demand fashion.
The reverse $k$-MIPS is therefore conducted online and is computationally-intensive task.
Now the question is how to efficiently obtain the reverse MIPS result for a given query.

A straightforward approach is to run a state-of-the-art exact MIPS algorithm \textit{for every vector in} $\mathbf{Q}$ and check whether or not $\mathbf{q} =$ arg\,max$_{\mathbf{p} \in \mathbf{P} \,\cup\, \{\mathbf{q}\}}\,\mathbf{u} \cdot \mathbf{p}$.
This approach obtains the exact result, but it incurs unnecessary computation.
The poor performance of this approach is derived from the following observations.
First, we do not need the MIPS result of $\mathbf{u}$ when $\mathbf{q}$ does not have the maximum inner product with $\mathbf{u}$.
Second, this approach certainly accesses all user vectors in $\mathbf{Q}$, although many of them do not contribute to the reverse MIPS result.
However, it is not trivial to skip evaluations of some user vectors without losing correctness.
Last, its theoretical cost is the same as the brute-force case, i.e., $O(nmd)$ time, where $n = |\mathbf{Q}|$ and $m = |\mathbf{P}|$, which is not appropriate for online computations.
These concerns pose challenges for solving the reverse MIPS problem efficiently.

\subsection{Contribution}
To address the above issues, we propose \textit{Simpfer}, a simple, fast, and exact algorithm for reverse MIPS.
The general idea of Simpfer is to efficiently solve the decision version of the MIPS problem.
Because the reverse MIPS of a query $\mathbf{q}$ requires a yes/no decision for each vector $\mathbf{u} \in \mathbf{Q}$, it is sufficient to know whether or not $\mathbf{q}$ can have the maximum inner product for $\mathbf{u}$.
Simpfer achieves this in $O(1)$ time in many cases by exploiting its index built in an offline phase.
This index furthermore supports a constant time filtering that prunes vectors in a batch if their answers are no.
We theoretically demonstrate that the time complexity of Simpfer is lower than $O(nmd)$.

The summary of our contributions is as follows:
\begin{itemize}
    \item   We address the problem of reverse $k$-MIPS.
            To our knowledge, this is the first work to study this problem.
    \item   We propose Simpfer as an exact solution to the reverse MIPS problem.
            Simpfer solves the decision version of the MIPS problem at both the group-level and the vector-level efficiently.
            Simpfer is surprisingly simple, but our analysis demonstrates that Simpfer theoretically outperforms a solution that employs a state-of-the-art exact MIPS algorithm.
    \item   We conduct extensive experiments on four real datasets, MovieLens, Netflix, Amazon, and Yahoo!.
            The results show that Simpfer is at least two orders magnitude faster than baselines.
    \item   Simpfer is easy to deploy: \textit{if recommender systems have user and item vector sets that are designed in the inner product space, they are ready to use Simpfer via our open source implementation\footnote{\url{https://github.com/amgt-d1/Simpfer}}.}
            This is because Simpfer is unsupervised and has only a single parameter (the maximum value of $k$) that is easy to tune and has no effect on the running time of online processing.
\end{itemize}
This paper is an error-corrected version of \cite{amagata2021reverse}.
We fixed some writing errors and minor bugs in our implementation, but our result is consistent with \cite{amagata2021reverse}.

\vspace{1.5mm}
\noindent
\textbf{Organization.}
The rest of this paper is organized as follows.
We formally define our problem in Section \ref{section_preliminary}.
We review related work in Section \ref{section_related-work}.
Our proposed algorithm is presented in Section \ref{section_simpfer}, and the experimental results are reported in Section \ref{section_experiment}.
Last, we conclude this paper in Section \ref{section_conclusion}.

\section{Problem Definition}   \label{section_preliminary}
Let $\mathbf{P}$ be a set of $d$-dimensional real-valued item vectors, and we assume that $d$ is high \cite{liu2019bandit,shrivastava2014asymmetric}.
Given a query vector, the maximum inner product search (MIPS) problem finds
\begin{equation}
    \mathbf{p^{*}} = \argmax_{\mathbf{p} \in \mathbf{P}}\,\mathbf{p}\cdot \mathbf{q}.    \nonumber
\end{equation}
The general version of the MIPS problem, i.e., the $k$-MIPS problem, is defined as follows:

\begin{defi}[\textsc{$k$-MIPS problem}]
Given a set of vectors $\mathbf{P}$, a query vector $\mathbf{q}$, and $k$, the $k$-MIPS problem finds $k$ vectors in $\mathbf{P}$ that have the highest inner products with $\mathbf{q}$.
\end{defi}

For a user (i.e., query), the $k$-MIPS problem can retrieve $k$ \textit{items} (e.g., vectors in $\mathbf{P}$) that the user would prefer.
Different from this, the reverse $k$-MIPS problem can retrieve \textit{a set of users} who would prefer a given item.
That is, in the reverse $k$-MIPS problem, a query can be an item, and \textit{this problem finds users attracted by the query item}.
Therefore, the reverse $k$-MIPS is effective for advertisement and market analysis, as described in Section \ref{section_introduction}.
We formally define this problem\footnote{Actually, the reverse top-k query (and its variant), a similar concept to the reverse $k$-MIPS problem, has been proposed in \cite{vlachou2010reverse, vlachou2011monochromatic, zhang2014reverse}.
It is important to note that these works do not suit recent recommender systems.
First, they assume that $d$ is low ($d$ is around 5), which is not probable in Matrix Factorization.
Second, they consider the Euclidean space, whereas inner product is a non-metric space.
Because the reverse top-k query processing algorithms are optimized for these assumptions, they cannot be employed in Matrix Factorization-based recommender systems and cannot solve (or be extended for) the reverse $k$-MIPS problem.}.

\begin{defi}[\textsc{Reverse $k$-MIPS problem}] \label{definition_rmips}
Given a query (item) vector $\mathbf{q}$, $k$, and two sets of vectors $\mathbf{Q}$ (set of user vectors) and $\mathbf{P}$ (set of item vectors), the reverse $k$-MIPS problem finds all vectors $\mathbf{u} \in \mathbf{Q}$ such that $\mathbf{q}$ is included in the $k$-MIPS result of $\mathbf{u}$ among $\mathbf{P} \cup \{\mathbf{q}\}$.
\end{defi}

\noindent
Note that $\mathbf{q}$ can be $\mathbf{q} \in \mathbf{P}$, as described in Example \ref{example_rmips}.
We use $n$ and $m$ to denote $|\mathbf{Q}|$ and $|\mathbf{P}|$, respectively.

Our only assumption is that there is a maximum $k$ that can be specified, denoted by $k_{max}$.
This is practical, because $k$ should be small, e.g., $k = 5$ \cite{jiang2020clustering} or $k = 10$ \cite{bachrach2014speeding}, to make applications effective.
(We explain how to deal with the case of $k > k_{max}$ in Section \ref{section_pre-processing}.)
This paper develops an exact solution to the new problem in Definition \ref{definition_rmips}.

\section{Related Work}  \label{section_related-work}
\underline{\textbf{Exact $k$-MIPS Algorithm.}}
The reverse $k$-MIPS problem can be solved exactly by conducting an exact $k$-MIPS algorithm for each user vector in $\mathbf{Q}$.
The first line of solution to the $k$-MIPS problem is a tree-index approach \cite{koenigstein2012efficient, ram2012maximum, curtin2013fast}.
For example, \cite{ram2012maximum} proposed a tree-based algorithm that processes $k$-MIPS not only for a single user vector but also for some user vectors in a batch.
Unfortunately, the performances of the tree-index algorithms degrade for large $d$ because of the curse of dimensionality.

LEMP \cite{teflioudi2015lemp,teflioudi2016exact} avoids this issue and significantly outperforms the tree-based algorithms.
LEMP uses several search algorithms according to the norm of each vector.
In addition, LEMP devises an early stop scheme of inner product computation.
During the computation of $\mathbf{u} \cdot \mathbf{q}$, LEMP computes an upper-bound of $\mathbf{u} \cdot \mathbf{q}$.
If this bound is lower than an intermediate $k$-th maximum inner product, $q$ cannot be in the final result, thus the inner product computation can be stopped.
LEMP is actually designed for the top-k inner product join problem: for each $\mathbf{u} \in \mathbf{Q}$, it finds the $k$-MIPS result of $\mathbf{u}$.
Therefore, LEMP can solve the reverse $k$-MIPS problem, but it is not efficient as demonstrated in Section \ref{section_experiment}.

FEXIPRO \cite{li2017fexipro} further improves the early stop of inner product computation of LEMP.
Specifically, FEXIPRO exploits singular value decomposition, integer approximation, and a transformation to positive values.
These techniques aim at obtaining a tighter upper-bound of $\mathbf{u} \cdot \mathbf{q}$ as early as possible.
\cite{li2017fexipro} reports that state-of-the-art tree-index algorithm \cite{ram2012maximum} is completely outperformed by FEXIPRO.
Maximus \cite{abuzaid2019index} takes hardware optimization into account.
It is, however, limited to specific CPUs, so we do not consider Maximus.
Note that LEMP and FEXIPRO are heuristic algorithms, and $O(nmd)$ time is required for the reverse $k$-MIPS problem.

\vspace{1.5mm}
\noindent
\underline{\textbf{Approximation $k$-MIPS Algorithm.}}
To solve the $k$-MIPS problem in sub-linear time by sacrificing correctness, many works proposed approximation $k$-MIPS algorithms.
There are several approaches to the approximation $k$-MIPS problem: sampling-based \cite{cohen1999approximating, liu2019bandit, yu2017greedy}, LSH-based \cite{huang2018accurate, neyshabur2015symmetric, shrivastava2014asymmetric, yan2018norm}, graph-based \cite{liu2020understanding, morozov2018non, zhou2019mobius}, and quantization approaches \cite{dai2020norm, guo2020accelerating}.
They have both strong and weak points.
For example, LSH-based algorithms enjoy a theoretical accuracy guarantee.
However, they are empirically slower than graph-based algorithms that have no theoretical performance guarantee.
Literature \cite{bachrach2014speeding} shows that the MIPS problem can be transformed into the Euclidean nearest neighbor search problem, but it still cannot provide the correct answer.
Besides, existing works that address the (reverse) nearest neighbor search problem assume low-dimensional data \cite{yang2015reverse} or consider approximation algorithms \cite{li2019approximate}.

Since this paper focuses on the exact result, these approximation $k$-MIPS algorithms cannot be utilized.
In addition, approximate answers may lose effectiveness of the reverse $k$-MIPS problem.
If applications cannot contain users, who are the answer of the $k$-MIPS problem, these users may lose chances of knowing the target item, which would reduce profits.
On the other hand, if applications contain users, who are \textit{not} the answer of the $k$-MIPS problem, as an approximate answer, they advertise the target item to users who are not interested in the item.
This also may lose future profits, because such users may stop receiving advertisements if they get those of non-interesting items.

\section{Proposed Algorithm}    \label{section_simpfer}
To efficiently solve the reverse MIPS problem, we propose Simpfer.
Its general idea is to efficiently solve the decision version of the $k$-MIPS problem.

\begin{defi}[\textsc{$k$-MIPS decision problem}]
Given a query $\mathbf{q}$, $k$, a user vector $\mathbf{u}$, and $\mathbf{P}$, this problem returns yes (no) if $\mathbf{q}$ is (not) included in the $k$-MIPS result of $\mathbf{u}$.
\end{defi}

\noindent
Notice that \textit{this problem does not require the complete $k$-MIPS result}.
We can terminate the $k$-MIPS of $\mathbf{u}$ whenever it is guaranteed that $\mathbf{q}$ is (not) included in the $k$-MIPS result.

To achieve this early termination efficiently, it is necessary to obtain a lower-bound and an upper-bound of the $k$-th highest inner product of $\mathbf{u}$.
Let $\phi$ and $\mu$ respectively be a lower-bound and an upper-bound of the $k$-th highest inner product of $\mathbf{u}$ on $\mathbf{P}$.
If $\phi \geq \mathbf{u} \cdot \mathbf{q}$, it is guaranteed that $\mathbf{q}$ does not have the $k$ highest inner product with $\mathbf{u}$.
Similarly, if $\mu \leq \mathbf{u} \cdot \mathbf{q}$, it is guaranteed that $\mathbf{q}$ has the $k$ highest inner product with $\mathbf{u}$.
This observation implies that we need to efficiently obtain $\phi$ and $\mu$.
Simpfer does pre-processing to enable it in an offline phase.
Besides, since $n = |\mathbf{Q}|$ is often large, accessing all user vectors is also time-consuming.
This requires a filtering technique that enables the pruning of user vectors that are not included in the reverse $k$-MIPS result \textit{in a batch}.
During the pre-processing, Simpfer arranges $\mathbf{Q}$ so that batch filtering is enabled.
Simpfer exploits the data structures built in the pre-processing phase to quickly solve the $k$-MIPS decision problem.

\subsection{Pre-processing} \label{section_pre-processing}
The objective of this pre-processing phase is to build data structures that support efficient computation of a lower-bound and an upper-bound of the $k$-th highest inner product for each $\mathbf{u_{i}} \in \mathbf{Q}$, for arbitrary queries.
We utilize Cauchy–Schwarz inequality for upper-bounding.
Hence we need the Euclidean norm $\|\mathbf{u_{i}}\|$ for each $\mathbf{u_{i}} \in \mathbf{Q}$.
To obtain a lower-bound of the $k$-th highest inner product, we need to access at least $k$ item vectors in $\mathbf{P}$.
The norm computation and lower-bound computation are independent of queries (as long as $k \leq k_{max}$), so they can be pre-computed.
In this phase, Simpfer builds the following array for each $\mathbf{u_{i}} \in \mathbf{Q}$.

\begin{defi}[\textsc{Lower-bound array}]
The lower-bound array $L_{i}$ of a user vector $\mathbf{u_{i}} \in \mathbf{Q}$ is an array whose $j$-th element, $L_{i}^{j}$, maintains a lower-bound of the $j$-th inner product of $\mathbf{u_{i}}$ on $\mathbf{P}$, and $|L_{i}| = k_{max}$.
\end{defi}

\noindent
Furthermore, to enable batch filtering, Simpfer builds a \textit{block}, which is defined below.

\begin{defi}[\textsc{Block}]    \label{definition_block}
A block $\mathbf{B}$ is a subset of $\mathbf{Q}$.
The set of vectors belonging to $\mathbf{B}$ is represented by $\mathbf{Q}(\mathbf{B})$.
Besides, we use $L(\mathbf{B})$ to represent the lower-bound array of this block, and
\begin{equation}
    L^{j}(\mathbf{B}) = \min_{\mathbf{u_{i}} \in \mathbf{Q}(\mathbf{B})} L_{i}^{j}  \label{equation_block}
\end{equation}
\end{defi}

\noindent
The block size $|\mathbf{Q}(\mathbf{B})|$ can be arbitrarily determined, and we set $|\mathbf{Q}(\mathbf{B})| = O(\log n)$ to avoid system parameter setting.

\begin{algorithm}[!t]
	\caption{\textsc{Pre-Processing of Simpfer}}
	\label{algo_pre-processing}
	\DontPrintSemicolon
    \KwIn {$\mathbf{Q}$, $\mathbf{P}$, and $k_{max}$}
        
    \For {each $\mathbf{u_{i}} \in \mathbf{Q}$} {
        Compute $\|\mathbf{u_{i}}\|$
    }
    \For {each $\mathbf{p_{j}} \in \mathbf{P}$} {
        Compute $\|\mathbf{p_{j}}\|$
    }
    Sort $\mathbf{Q}$ and $\mathbf{P}$ in descending order of norm size\;
    $\mathbf{P'} \leftarrow$ the first $O(k_{max})$ vectors in $\mathbf{P}$\;
    \For {each $\mathbf{u_{i}} \in \mathbf{Q}$}{
        $\mathbf{R} \leftarrow k_{max}$ vectors $\mathbf{p} \in \mathbf{P'}$ that maximize $\mathbf{u_{i}} \cdot \mathbf{p}$\;
        \For {$j = 1$ to $k_{max}$} {
            $L_{i}^{j} \leftarrow \mathbf{u_{i}} \cdot \mathbf{p}$, where $\mathbf{p}$ provides the $j$-th highest inner product with $\mathbf{u_{i}}$ in $\mathbf{R}$
        }
    }
    $\mathcal{B} \leftarrow \varnothing$, $\mathbf{B} \leftarrow$ a new block\;
    \For {each $\mathbf{u_{i}} \in \mathbf{Q}$} {
        $\mathbf{Q}(\mathbf{B}) \leftarrow \mathbf{Q}(\mathbf{B}) \cup \{\mathbf{u_{i}}\}$\;
        \For {$j = 1$ to $k_{max}$}{
            $L^{j}(\mathbf{B}) \leftarrow \min\{L^{j}(\mathbf{B}), L_{i}^{j}\}$
        }
        \If {$|\mathbf{Q}(\mathbf{B})| = O(\log n)$} {
            $\mathcal{B} \leftarrow \mathcal{B} \cup \{\mathbf{B}\}$\;
            $\mathbf{B} \leftarrow$ a new block
        }
    }
\end{algorithm}

\vspace{1.5mm}
\noindent
\underline{\textbf{Pre-processing algorithm.}}
Algorithm \ref{algo_pre-processing} describes the pre-processing algorithm of Simpfer.

\vspace{1.5mm}
\noindent
(1) Norm computation: First, for each $\mathbf{u} \in \mathbf{Q}$ and $\mathbf{p} \in \mathbf{P}$, its norm is computed.
Then, $\mathbf{Q}$ and $\mathbf{P}$ are sorted in descending order of norm.

\vspace{1.5mm}
\noindent
(2) Lower-bound array building: Let $\mathbf{P'}$ be the set of the first $O(k_{max})$ vectors in $\mathbf{P}$.
For each $\mathbf{u_{i}} \in \mathbf{Q}$, $L_{i}$ is built by using $\mathbf{P'}$.
That is, $L_{i}^{j} = \mathbf{u_{i}} \cdot \mathbf{p}$, where $\mathbf{p} \in \mathbf{P'}$ yields the $j$-th highest inner product for $\mathbf{u} \in \mathbf{P'}$.
The behind idea of using the first $O(k_{max})$ item vectors in $\mathbf{P}$ is that vectors with large norms tend to provide large inner products \cite{liu2020understanding}.
This means that we can obtain a tight lower-bound at a lightweight cost.

\vspace{1.5mm}
\noindent
(3) Block building: After that, blocks are built, so that user vectors in a block keep the order and each block is disjoint.
Given a new block $\mathbf{B}$, we insert user vectors $\mathbf{u_{i}} \in \mathbf{Q}$ into $\mathbf{Q}(\mathbf{B})$ in sequence while updating $L^{j}(\mathbf{B})$, until we have $|\mathbf{Q}(\mathbf{B})| = O(\log n)$.
When $|\mathbf{Q}(\mathbf{B})| = O(\log n)$, we insert $\mathbf{B}$ into a set of blocks $\mathcal{B}$, and make a new block.

\begin{figure}[!t]
	\centering
    \includegraphics[width=0.95\linewidth]{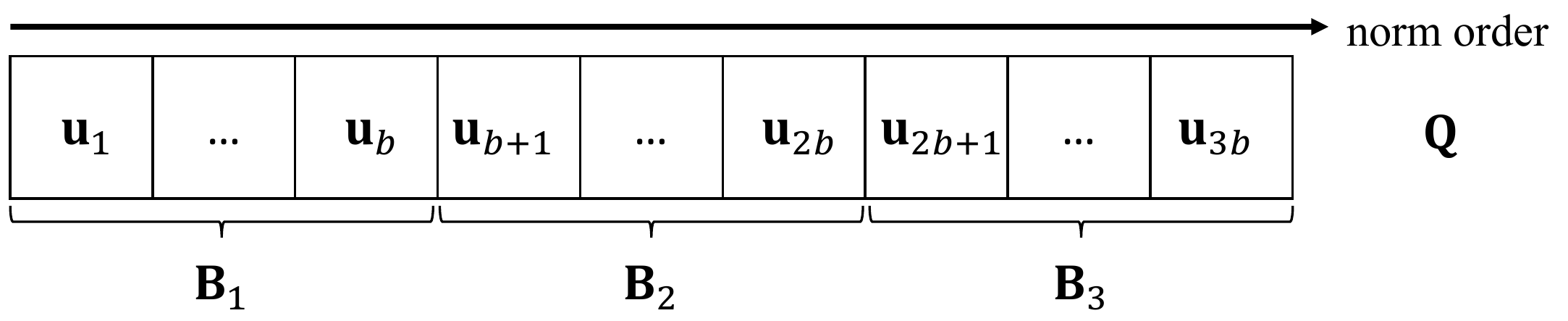}
	\caption{Example of block building.}
    \label{figure_block}
    \Description[Block structure]{Our block structure for a set $\mathbf{Q}$ of user vectors, where they are sorted in descending order of norm. Every block is disjoint, and each user vector is assigned to a unique block.}
\end{figure}

\begin{exa}
Figure \ref{figure_block} illustrates an example of block building.
For ease of presentation, we use $b$ as a block size and $n = 3b$.
For example, $\mathbf{Q}(\mathbf{B_{1}}) = \{\mathbf{u_{1}},...,\mathbf{u_{b}}\}$, and $\|\mathbf{u_{1}}\| \geq ...\geq \|\mathbf{u_{b}}\|$.
\end{exa}

Generally, this pre-processing is done only once.
An exception is the case where a query with $k > k_{max}$ is specified.
In this case, Simpfer re-builds the data structures then processes the query.
This is actually much faster than the baselines, as shown in Section \ref{section_experiment-offline}.

\vspace{1.5mm}
\noindent
\underline{\textbf{Analysis.}}
We here prove that the time complexity of this pre-processing is reasonable.
Without loss of generality, we assume $n \geq m$, because this is a usual case for many real datasets, as the ones we use in Section \ref{section_experiment}.

\begin{theo}    \label{theorem_pre-time}
Algorithm \ref{algo_pre-processing} requires $O(n(d + \log n))$ time.
\end{theo}

\noindent
\textsc{Proof.}
The norm computation requires $O((n+m)d) = O(nd)$ time, and sorting requires $O(n\log n)$ time.
The building of lower-bound arrays needs $O(n \times k_{max})$ time, since $O(|\mathbf{P'}|) = O(k_{max})$.
Because $k_{max} = O(1)$, $O(n \times k_{max}) = O(n)$.
The block building also requires $O(n \times k_{max}) = O(n)$ time.
In total, this pre-processing requires $O(n(d+\log n))$ time.   \wsq

\vspace{1.5mm}
\noindent
The space complexity of Simpfer is also reasonable.

\begin{theo}
The space complexity of the index is $O(n)$.
\end{theo}

\noindent
\textsc{Proof.}
The space of the lower-bound arrays of user vectors is $O(\sum_{n}|L_{i}|) = O(n)$, since $O(|L_{i}|) = O(1)$.
Blocks are disjoint, and the space of the lower-bound array of a block is also $O(1)$.
We hence have $O(\frac{n}{\log n})$ lower-bound arrays of blocks.
Now this theorem is clear.  \wsq

\subsection{Upper- and Lower-bounding for the $k$-MIPS Decision Problem}    \label{section_bounding}
Before we present the details of Simpfer, we introduce our techniques that can quickly answer the $k$-MIPS decision problem for a given query $\mathbf{q}$.
Recall that $\mathbf{Q}$ and $\mathbf{P}$ are sorted in descending order of norm.
Without loss of generality, we assume that $\|\mathbf{u_{i}}\| \geq \|\mathbf{u_{i+1}}\|$ for each $i \in [1,n-1]$ and $\|\mathbf{p_{j}}\| \geq \|\mathbf{p_{j+1}}\|$ for each $j \in [1,m-1]$, for ease of presentation.

Given a query $\mathbf{q}$ and a user vector $\mathbf{u_{i}} \in \mathbf{Q}$, we have $\mathbf{u_{i}} \cdot \mathbf{q}$.
Although our data structures are simple, they provide effective and ``light-weight'' filters.
Specifically, we can quickly answer the $k$-MIPS decision problem on $\mathbf{q}$ through the following observations\footnote{Existing algorithms for top-k retrieval, e.g., \cite{ding2011faster, fontoura2011evaluation}, use similar (but different) bounding techniques.
They use a bound (e.g., obtained by a block) to \textit{early stop} linear scans.
On the other hand, our bounding is designed to \textit{avoid} linear scans and to filer multiple user vectors in a batch.}.

\begin{lemm}    \label{lemma_vector-no}
If $\mathbf{u_{i}} \cdot \mathbf{q} \leq L_{i}^{k}$, it is guaranteed that $\mathbf{q}$ is not included in the $k$-MIPS result of $\mathbf{u_{i}}$.
\end{lemm}

\noindent
\textsc{Proof.}
Let $\mathbf{p}$ be the vector in $\mathbf{P}$ such that $\mathbf{u_{i}} \cdot \mathbf{p}$ is the $k$-th highest inner product in $\mathbf{P}$.
The fact that $L_{i}^{k} \leq \mathbf{u_{i}} \cdot \mathbf{p}$ immediately derives this lemma.    \wsq

\vspace{1.5mm}
\noindent
It is important to see that the above lemma provides ``no'' as the answer to the $k$-MIPS decision problem on $\mathbf{q}$ in $O(1)$ time (after computing $\mathbf{u_{i}} \cdot \mathbf{q}$).
The next lemma deals with the ``yes'' case in $O(1)$ time.

\begin{lemm}    \label{lemma_vector-yes}
If $\mathbf{u_{i}} \cdot \mathbf{q} \geq \|\mathbf{u_{i}}\|\|\mathbf{p_{k}}\|$, it is guaranteed that $\mathbf{q}$ is included in the $k$-MIPS result of $\mathbf{u_{i}}$.
\end{lemm}

\noindent
\textsc{Proof.}
From Cauchy–Schwarz inequality, we have $\mathbf{u_{i}} \cdot \mathbf{p_{j}} \leq \|\mathbf{u_{i}}\|\|\mathbf{p_{j}}\|$.
Since $\|\mathbf{p_{k}}\|$ is the $k$-th highest norm in $\mathbf{P}$, $\mathbf{u_{i}} \cdot \mathbf{p} \leq \|\mathbf{u_{i}}\|\|\mathbf{p_{k}}\|$, where $\mathbf{p}$ is defined in the proof of Lemma \ref{lemma_vector-no}.
That is, $\|\mathbf{u_{i}}\|\|\mathbf{p_{k}}\|$ is an upper-bound of $\mathbf{u_{i}} \cdot \mathbf{p}$.
Now it is clear that $\mathbf{q}$ has $\mathbf{u_{i}} \cdot \mathbf{q} \geq \mathbf{u_{i}} \cdot \mathbf{p}$ if $\mathbf{u_{i}} \cdot \mathbf{q} \geq \|\mathbf{u_{i}}\|\|\mathbf{p_{k}}\|$. \wsq

\vspace{1.5mm}
We next introduce a technique that yields ``no'' as the answer for \textit{all user vectors} in a block $\mathbf{B}$ in $O(1)$ time.

\begin{lemm}    \label{lemma_block}
Given a block $\mathbf{B}$, let $\mathbf{u_{i}}$ be the first vector in $\mathbf{Q}(\mathbf{B})$.
If $\|\mathbf{u_{i}}\|\|\mathbf{q}\| \leq L^{k}(\mathbf{B})$, for all $\mathbf{u_{j}} \in \mathbf{Q}(\mathbf{B})$, it is guaranteed that $\mathbf{q}$ is not included in the $k$-MIPS result of $\mathbf{u_{j}}$.
\end{lemm}

\noindent
\textsc{Proof.}
From Cauchy–Schwarz inequality, $\|\mathbf{u_{i}}\|\|\mathbf{q}\|$ is an upper-bound of $\mathbf{u_{j}} \cdot \mathbf{q}$ for all $\mathbf{u_{j}} \in \mathbf{Q}(\mathbf{B})$, since $\mathbf{Q}(\mathbf{B}) = \{\mathbf{u_{i}},\mathbf{u_{i+1}},...\}$. 
We have $L^{k}(\mathbf{B}) \leq L_{j}^{k}$ for all $\mathbf{u_{j}} \in \mathbf{Q}(\mathbf{B})$, from Equation (\ref{equation_block}).
Therefore, if $\|\mathbf{u_{i}}\|\|\mathbf{q}\| \leq L^{k}(\mathbf{B})$, $\mathbf{u_{j}} \cdot \mathbf{q}$ cannot be the $k$ highest inner product. \wsq

\vspace{1.5mm}
If a user vector $\mathbf{u_{i}}$ cannot obtain a yes/no answer from Lemmas \ref{lemma_vector-no}--\ref{lemma_block}, Simpfer uses a linear scan of $\mathbf{P}$ to obtain the answer.
Let $\tau$ be a threshold, i.e., an intermediate $k$-th highest inner product for $\mathbf{u}$ during the linear scan.
By using the following corollaries, Simpfer can obtain the correct answer and early terminate the linear scan.

\begin{coro}    \label{corollary_yes}
Assume that $\mathbf{q}$ is included in an intermediate result of the $k$-MIPS of $\mathbf{u_{i}}$ and we now evaluate $\mathbf{p_{j}} \in \mathbf{P}$.
If $\mathbf{u_{i}} \cdot \mathbf{q} \geq \|\mathbf{u_{i}}\|\|\mathbf{p_{j}}\|$, it is guaranteed that $\mathbf{q}$ is included in the final result of the $k$-MIPS of $\mathbf{u_{i}}$.
\end{coro}

\noindent
\textsc{Proof.}
Trivially, we have $j \geq k$.
Besides, $\|\mathbf{u_{i}}\|\|\mathbf{p_{j}}\| \geq \mathbf{u_{i}} \cdot \mathbf{p_{l}}$ for all $k \leq l \leq m$, because $\mathbf{P}$ is sorted.
This corollary is hence true. \wsq

\vspace{1.5mm}
\noindent
From this corollary, we also have:

\begin{coro}    \label{corollary_no}
When we have $\tau > \mathbf{u_{i}} \cdot \mathbf{q}$, it is guaranteed that $\mathbf{q}$ is not included in the final result of the $k$-MIPS of $\mathbf{u_{i}}$.
\end{coro}

\noindent
Algorithm \ref{algo_linear-scan} summarizes the linear scan that incorporates Corollaries \ref{corollary_yes}--\ref{corollary_no}.

\subsection{The Algorithm}
Now we are ready to present Simpfer.
Algorithm \ref{algo_simpfer} details it.
To start with, Simpfer computes $\|\mathbf{q}\|$.
Given a block $\mathbf{B} \in \mathcal{B}$, Simpfer tests Lemma \ref{lemma_block} (line \ref{algo_simpfer_lemma-block}).
If the user vectors in $\mathbf{Q}(\mathbf{B})$ may have yes as an answer, for each $\mathbf{u_{i}} \in \mathbf{Q}(\mathbf{B})$, Simpfer does the following.
(Otherwise, all user vectors in $\mathbf{Q}(\mathbf{B})$ are ignored.)
First, it computes $\mathbf{u_{i}} \cdot \mathbf{q}$, then tests Lemma \ref{lemma_vector-no} (line \ref{algo_simpfer_lemma-vector-no}).
If $\mathbf{u_{i}}$ cannot have the answer from this lemma, Simpfer tests Lemma \ref{lemma_vector-yes}.
Simpfer inserts $\mathbf{u_{i}}$ into the result set $\mathbf{Q_{r}}$ if $\mathbf{u_{i}} \cdot \mathbf{q} \geq \|\mathbf{u_{i}}\|\|\mathbf{p_{k}}\|$.
Otherwise, Simpfer conducts \textsc{Linear-scan}$(\mathbf{u_{i}})$ (Algorithm \ref{algo_linear-scan}).
If \textsc{Linear-scan}$(\mathbf{u_{i}})$ returns 1 (yes), $\mathbf{u_{i}}$ is inserted into $\mathbf{Q_{r}}$.
The above operations are repeated for each $\mathbf{B} \in \mathcal{B}$.
Finally, Simpfer returns the result set $\mathbf{Q_{r}}$.

The correctness of Simpfer is obvious, because it conducts \textsc{Linear-scan}$(\cdot)$ for all vectors that cannot have yes/no answers from Lemmas \ref{lemma_vector-no}--\ref{lemma_block}.
Besides, Simpfer accesses blocks sequentially, so it is easy to parallelize by using multicore.
Simpfer hence further accelerates the processing of reverse $k$-MIPS, see Section \ref{section_experiment-multicore}.

\begin{algorithm}[!t]
	\caption{\textsc{Linear-scan$(u)$}}
	\label{algo_linear-scan}
	\DontPrintSemicolon
    \KwIn {$\mathbf{u} \in \mathbf{Q}$, $\mathbf{P}$, $\mathbf{q}$, and $k$}
    
    $I \leftarrow \{\mathbf{u} \cdot \mathbf{q}\}$, $\tau \leftarrow 0$\;
    \For {each $\mathbf{p_{i}} \in \mathbf{P}$} {
        \If {$\mathbf{u} \cdot \mathbf{q} \geq \|\mathbf{u}\|\|\mathbf{p_{i}}\|$} {
            \textbf{return} 1 (yes)
        }
        $\gamma \leftarrow \mathbf{u} \cdot \mathbf{p_{i}}$\;
        \If {$\gamma > \tau$} {
            $I \leftarrow I \cup \{\gamma\}$\;
            \If {$|I| > k$} {
                Delete the $(k+1)$-th inner product from $I$\;
                $\tau \leftarrow$ the $k$-th inner product in $I$
            }
            \If {$\tau > \mathbf{u} \cdot \mathbf{q}$} {
                \textbf{return} 0 (no)
            }
        }
    }
\end{algorithm}
\begin{algorithm}[!t]
	\caption{\textsc{Simpfer}}
	\label{algo_simpfer}
	\DontPrintSemicolon
    \KwIn {$\mathbf{Q}$, $\mathbf{P}$, $\mathbf{q}$, $k$, and $\mathcal{B}$}
    $\mathbf{Q_{r}} \leftarrow \varnothing$, Compute $\|\mathbf{q}\|$\;
    \For {each $\mathbf{B} \in \mathcal{B}$} {
        $\mathbf{u} \leftarrow$ the first user vector in $\mathbf{Q}(\mathbf{B})$\;
        \If {$\|\mathbf{u}\|\|\mathbf{q}\| > L^{k}(\mathbf{B})$} {                  \label{algo_simpfer_lemma-block}
            \For {each $\mathbf{u_{i}} \in \mathbf{Q}(\mathbf{B})$} {
                $\gamma \leftarrow \mathbf{u_{i}} \cdot \mathbf{q}$\;
                \If {$\gamma > L_{i}^{k}$} {                                        \label{algo_simpfer_lemma-vector-no}
                    \eIf {$\|\mathbf{u_{i}}\|\|\mathbf{p_{k}}\| > \gamma$} {        \label{algo_simpfer_lemma-vector-yes}
                        $f \leftarrow$ \textsc{Linear-Scan$(\mathbf{u_{i}})$}\;
                        \If {$f = 1$} {
                            $\mathbf{Q_{r}} \leftarrow \mathbf{Q_{r}} \cup \{\mathbf{u_{i}}\}$
                        }
                    }{
                        $\mathbf{Q_{r}} \leftarrow \mathbf{Q_{r}} \cup \{\mathbf{u_{i}}\}$
                    }
                }
            }
        }
    }
    \textbf{return} $\mathbf{Q_{r}}$
\end{algorithm}

\subsection{Complexity Analysis}
We theoretically demonstrate the efficiency of Simpfer.
Specifically, we have:

\begin{theo}    \label{theorem_time}
Let $\alpha$ be the pruning ratio ($0 \leq \alpha \leq 1$) of blocks in $\mathcal{B}$.
Furthermore, let $m'$ be the average number of item vectors accessed in \textsc{Linear-scan}$(\cdot)$.
The time complexity of Simpfer is $O((1-\alpha)nm'd)$.
\end{theo}

\noindent
\textsc{Proof.}
Simpfer accesses all blocks in $\mathcal{B}$, and $|\mathcal{B}| = O(\frac{n}{\log n})$.
Assume that a block $\mathbf{B} \in \mathcal{B}$ is not pruned by Lemma \ref{lemma_block}.
Simpfer accesses all user vectors in $\mathbf{Q}(\mathbf{B})$, so the total number of such user vectors is $(1 - \alpha) \times O(\frac{n}{\log n}) \times O(\log n) = O((1 - \alpha)n)$.
For these vectors, Simpfer computes inner products with $\mathbf{q}$.
The evaluation cost of Lemmas \ref{lemma_vector-no} and \ref{lemma_vector-yes} for these user vectors is thus $O((1 - \alpha)nd)$.
The worst cost of \textsc{Linear-scan}$(\cdot)$ for vectors that cannot obtain the answer from these lemmas is $O((1-\alpha)nm'd)$.
Now the time complexity of Simpfer is
\begin{align}
    O(\frac{n}{\log n} + (1 - \alpha)nd + (1 - \alpha)nm'd) &= O(\frac{n}{\log n} + (1 - \alpha)nm'd)   \label{equation_time}   \\
                                                            &= O((1 - \alpha)nm'd)  \nonumber
\end{align}
Consequently, this theorem holds.   \wsq

\vspace{1.5mm}
\noindent
\underline{\textbf{Remark.}}
There are two main observations in Theorem \ref{theorem_time}.
First, because we practically have $m' < m$ and $\alpha > 0$, Simpfer outperforms a $k$-MIPS-based solution that incurs $O(nmd)$ time.
(Our experimental results show that $m' = O(k)$ in practice.)
The second observation is obtained from Equation (\ref{equation_time}), which implies the effectiveness of blocks.
If Simpfer does not build blocks, we have to evaluate Lemma \ref{lemma_vector-no} \textit{for all} $\mathbf{u} \in \mathbf{Q}$.
Equation (\ref{equation_time}) suggests that the blocks theoretically avoids this.

\section{Experiment}    \label{section_experiment}
This section reports our experimental results.
All experiments were conducted on a Ubuntu 18.04 LTS machine with a 12-core 3.0GHz Intel Xeon E5-2687w v4 processor and 512GB RAM.

\subsection{Setting}
\noindent
\underline{\textbf{Datasets.}}
We used four popular real datasets: MovieLens\footnote{\url{https://grouplens.org/datasets/movielens/}}, Netflix, Amazon\footnote{\url{https://jmcauley.ucsd.edu/data/amazon/}}, and Yahoo!\footnote{\url{https://webscope.sandbox.yahoo.com/}}.
The user and item vectors of these datasets were obtained by the Matrix Factorization in \cite{chin2016libmf}.
These are 50-dimensional vectors (the dimensionality setting is the same as \cite{li2017fexipro, teflioudi2015lemp}\footnote{As our theoretical analysis shows, the time of Simpfer is trivially proportional to $d$, thus its empirical impact is omitted.}).
The other statistics is shown in Table \ref{table_statistics}.
We randomly chose 1,000 vectors as query vectors from $\mathbf{P}$.

\begin{table}[!h]
    \centering
	\caption{Dataset statistics}
	\label{table_statistics}
	\Description[Cardinality of user and item vector sets]{Cardinality of user and item vector sets for MovieLens, Netflix, Amazon, and Yahoo!.}
	\begin{tabular}{c||c|c|c|c} \hline
                        & MovieLens	& Netflix   & Amazon    & Yahoo!    \\ \hline \hline
        $|\mathbf{Q}|$	& 138,493   & 480,189   & 1,948,882 & 2,088,620 \\ \hline
        $|\mathbf{P}|$	& 26,744	& 17,770	& 98,211	& 200,941	\\ \hline
	\end{tabular}
\end{table}

\vspace{1.5mm}
\noindent
\underline{\textbf{Evaluated algorithms.}}
We evaluated the following three algorithms.
\begin{itemize}
    \item   LEMP \cite{teflioudi2015lemp}: the state-of-the-art \textit{all}-$k$-MIPS algorithm.
            LEMP originally does $k$-MIPS for all user vectors in $\mathbf{Q}$.
    \item   FEXIPRO \cite{li2017fexipro}: the state-of-the-art $k$-MIPS algorithm.
            We simply ran FEXIPRO for each $\mathbf{u} \in \mathbf{Q}$.
    \item   Simpfer: the algorithm proposed in this paper.
            We set $k_{max} = 25$.
\end{itemize}
These algorithms were implemented in C++ and compiled by g++ 7.5.0 with -O3 flag.
We used OpenMP for multicore processing.
These algorithms return the exact result, so we measured their running time.

Note that \cite{li2017fexipro, teflioudi2015lemp} have demonstrated that the other exact MIPS algorithms are outperformed by LEMP and FEXIPRO, so we did not use them as competitors.
(Recall that this paper focuses on the exact answer, thus approximation algorithms are not appropriate for competitors, see Section \ref{section_related-work}.)
In addition, LEMP and FEXIPRO also have a pre-processing (offline) phase.
We did not include the offline time as the running time.

\subsection{Result 1: Effectiveness of blocks}
We first clarify the effectiveness of blocks employed in Simpfer.
To show this, we compare Simpfer with Simpfer without blocks (which does not evaluate line \ref{algo_simpfer_lemma-block} of Algorithm \ref{algo_simpfer}).
We set $k = 10$.

On MovieLens, Netflix, Amazon, and Yahoo!, Simpfer (Simpfer without blocks) takes 10.3 (22.0), 58.6 (100.8), 117.6 (446.2), and 1481.2 (1586.2) [msec], respectively.
This result demonstrates that, although the speed-up ratio is affected by data distributions, blocks surely yield speed-up.

\subsection{Result 2: Impact of $k$}
\begin{figure*}[t]
	\begin{center}
        \subfigure[MovieLens (Running time)]{%
		\includegraphics[width=0.24\linewidth]{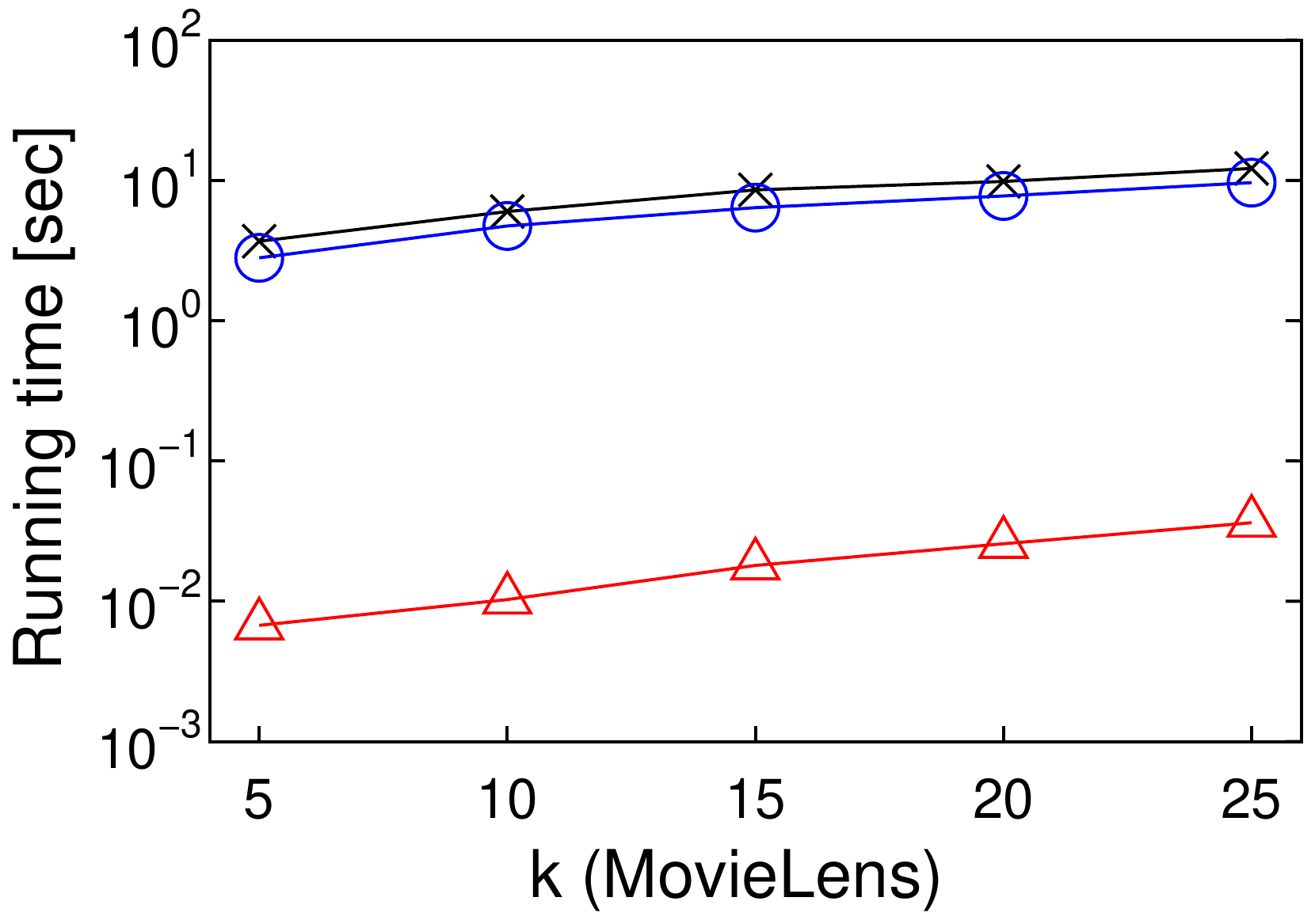}  \label{figure_movielens-k}}
        \subfigure[Netflix (Running time)]{%
		\includegraphics[width=0.24\linewidth]{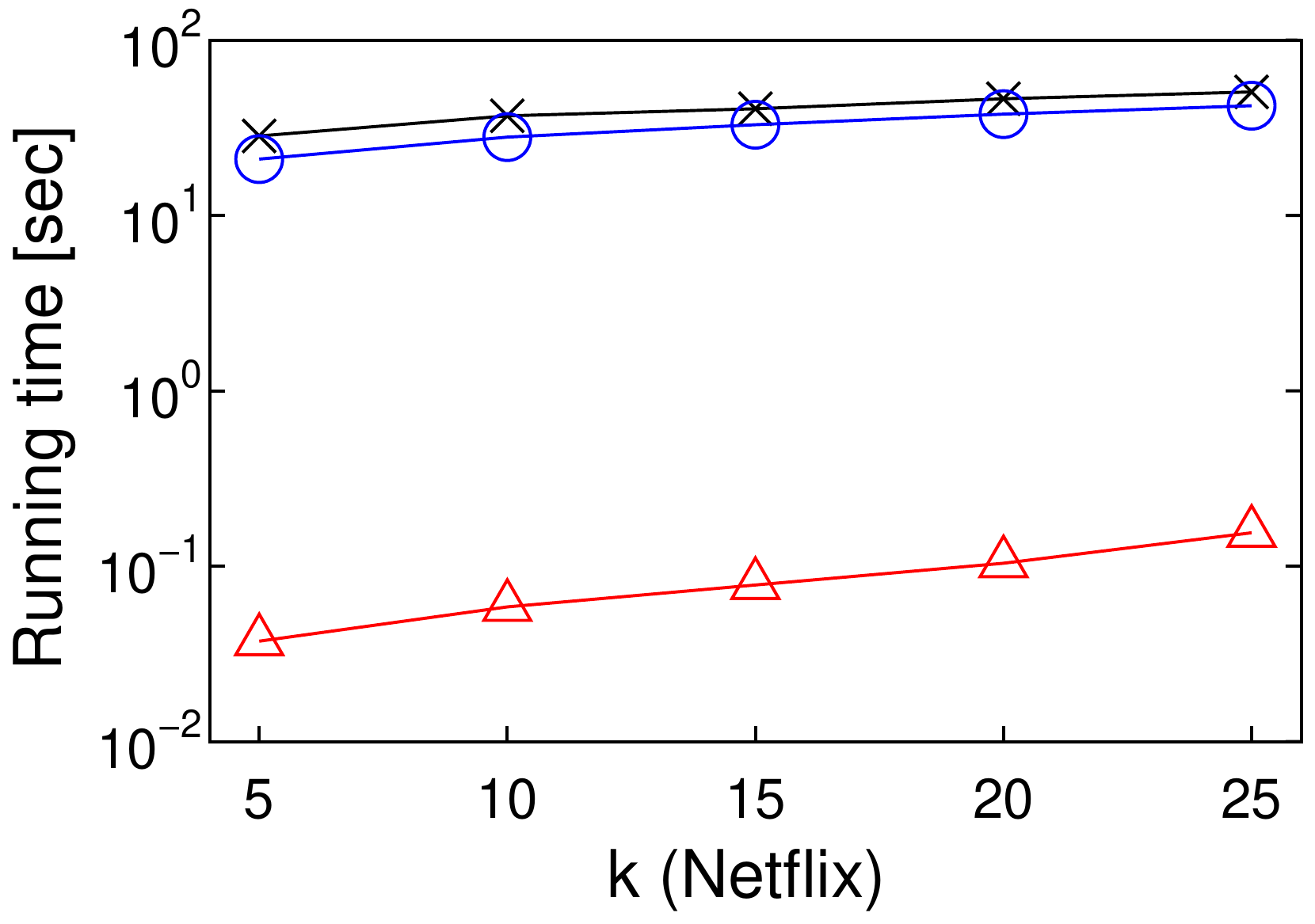}	\label{figure_netflix-k}}
		\subfigure[Amazon (Running time)]{%
		\includegraphics[width=0.24\linewidth]{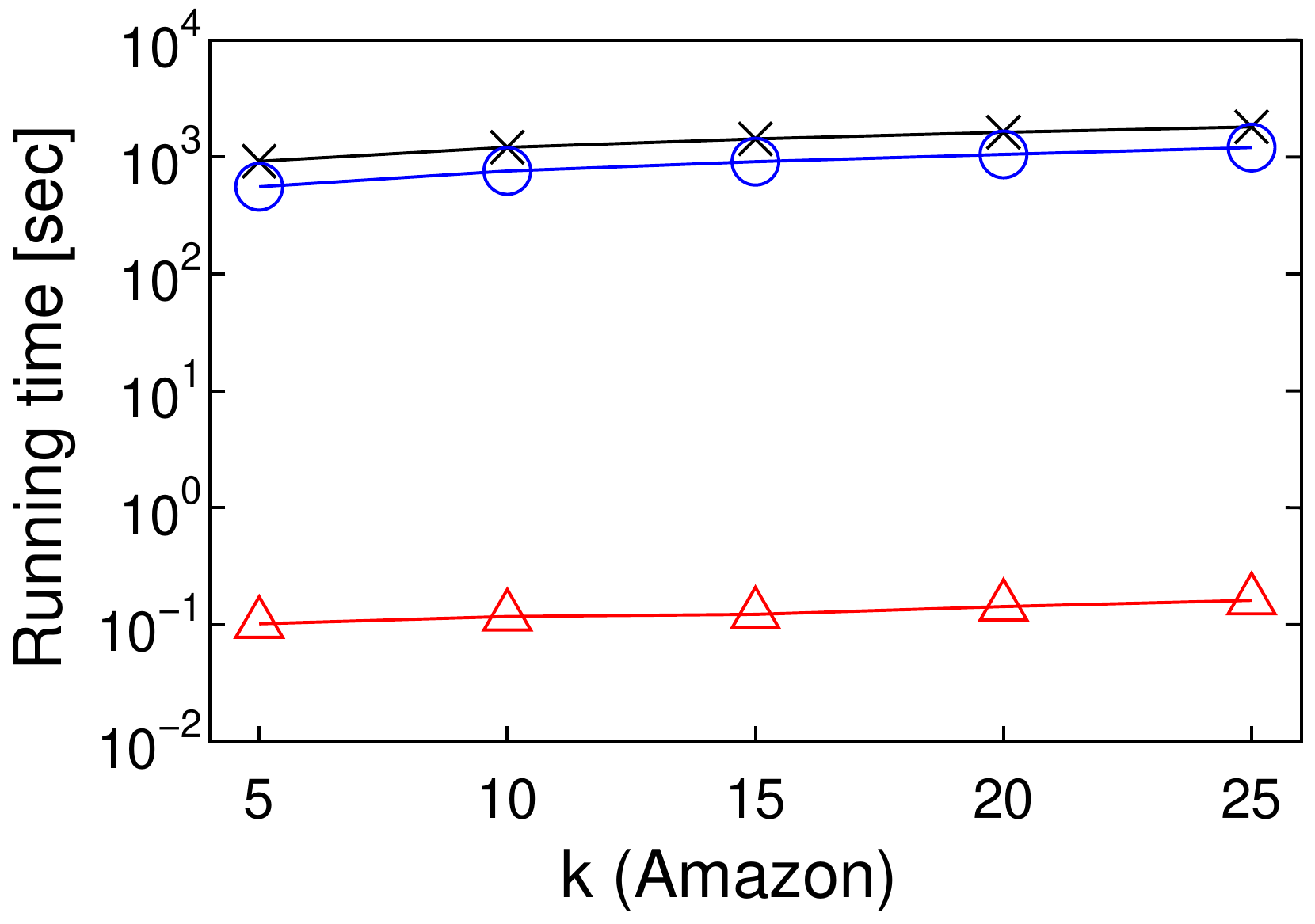}	    \label{figure_amazon-k}}
        \subfigure[Yahoo! (Running time)]{%
		\includegraphics[width=0.24\linewidth]{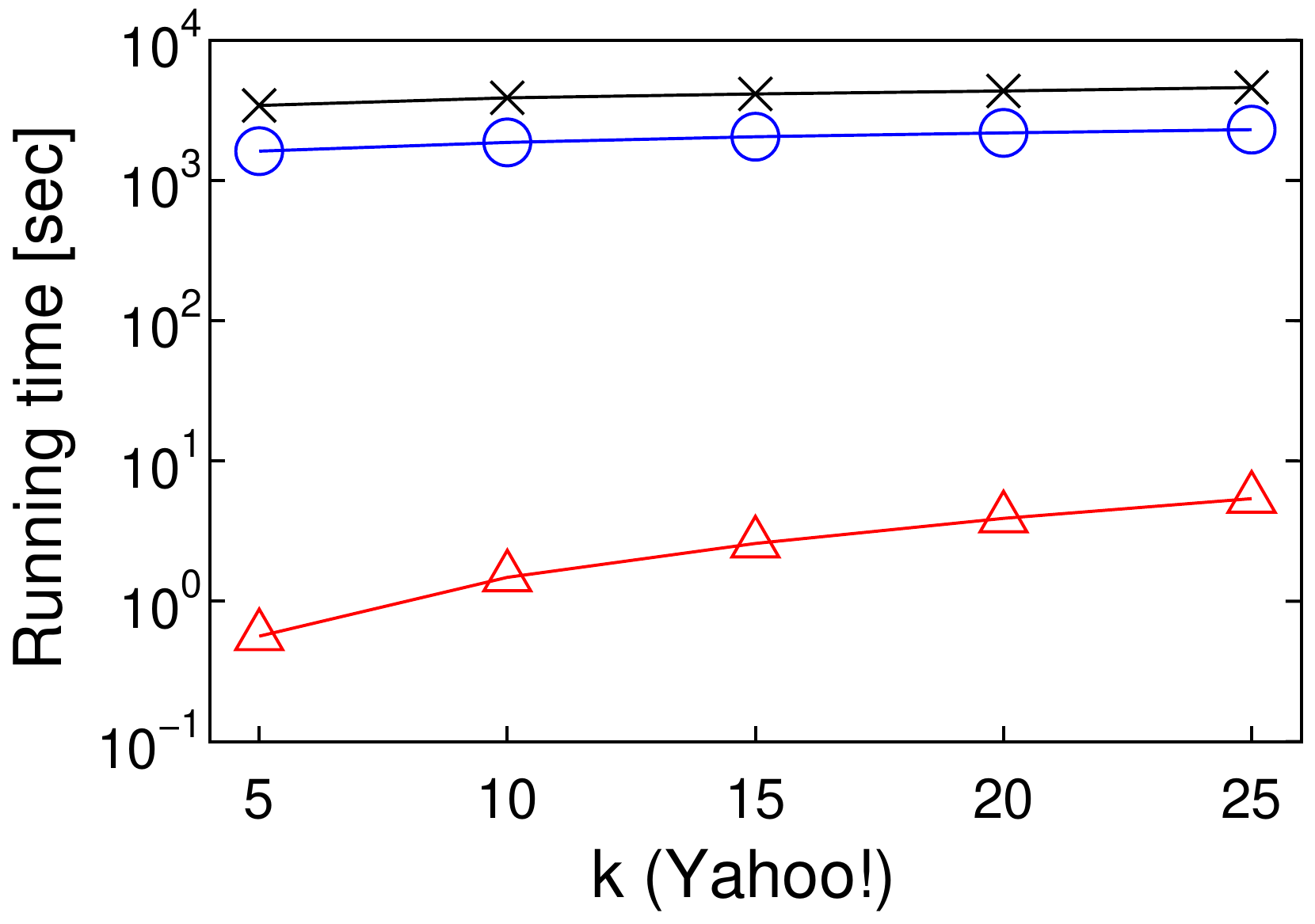}  	\label{figure_yahoo-k}}
		\subfigure[MovieLens (\#ip computations)]{%
		\includegraphics[width=0.24\linewidth]{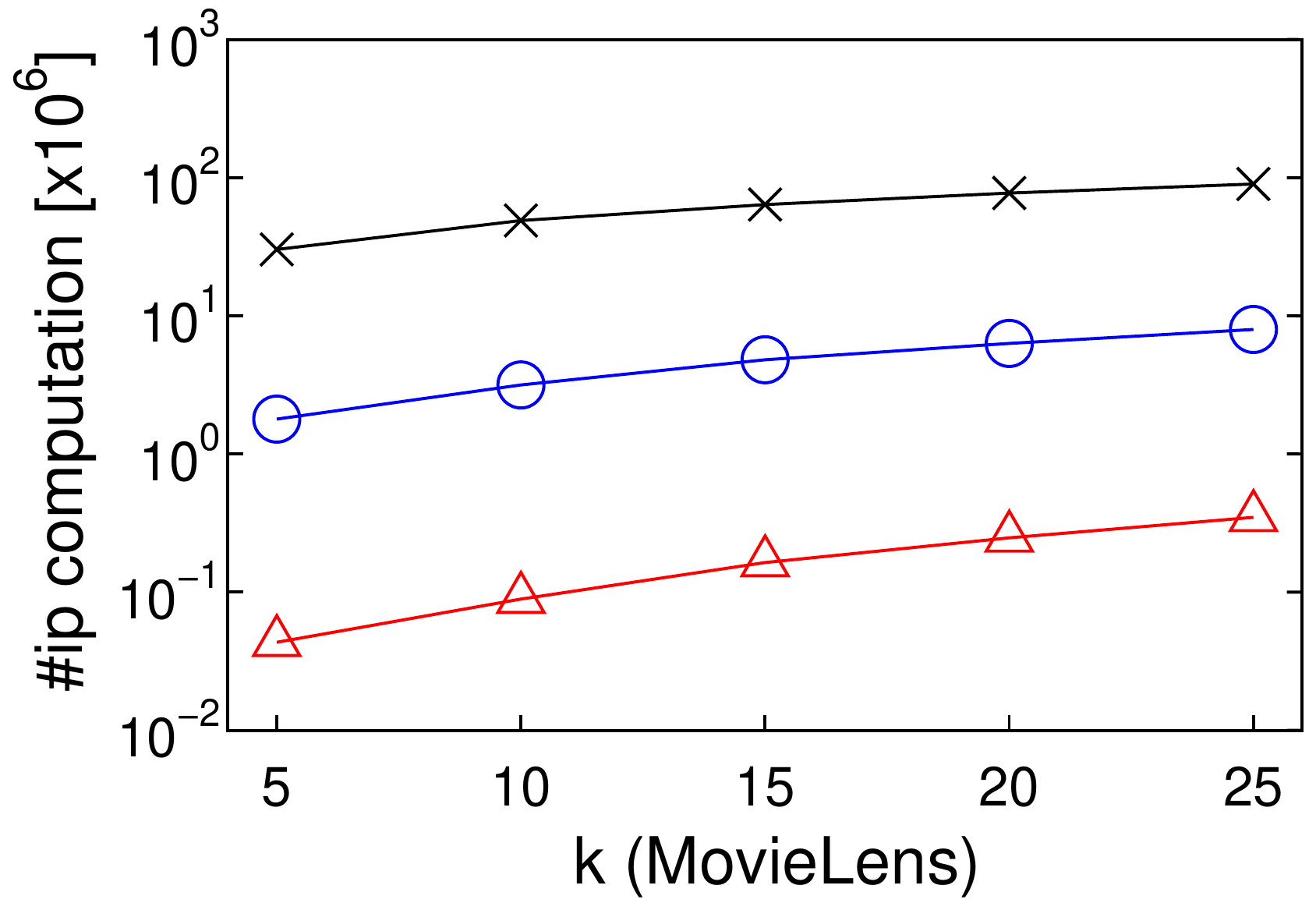}  \label{figure_movielens-k_ip}}
        \subfigure[Netflix (\#ip computations)]{%
		\includegraphics[width=0.24\linewidth]{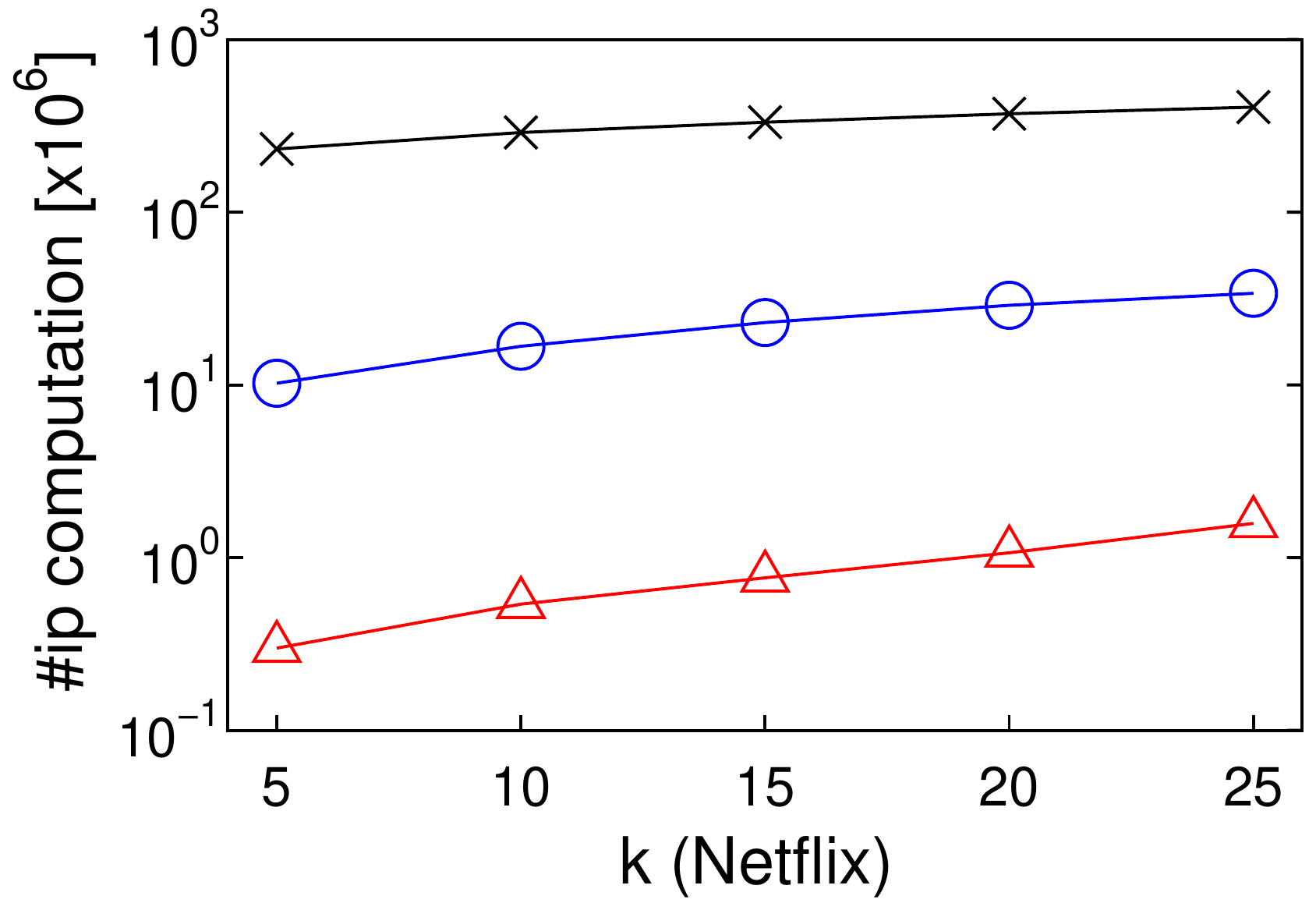}	    \label{figure_netflix-k_ip}}
		\subfigure[Amazon (\#ip computations)]{%
		\includegraphics[width=0.24\linewidth]{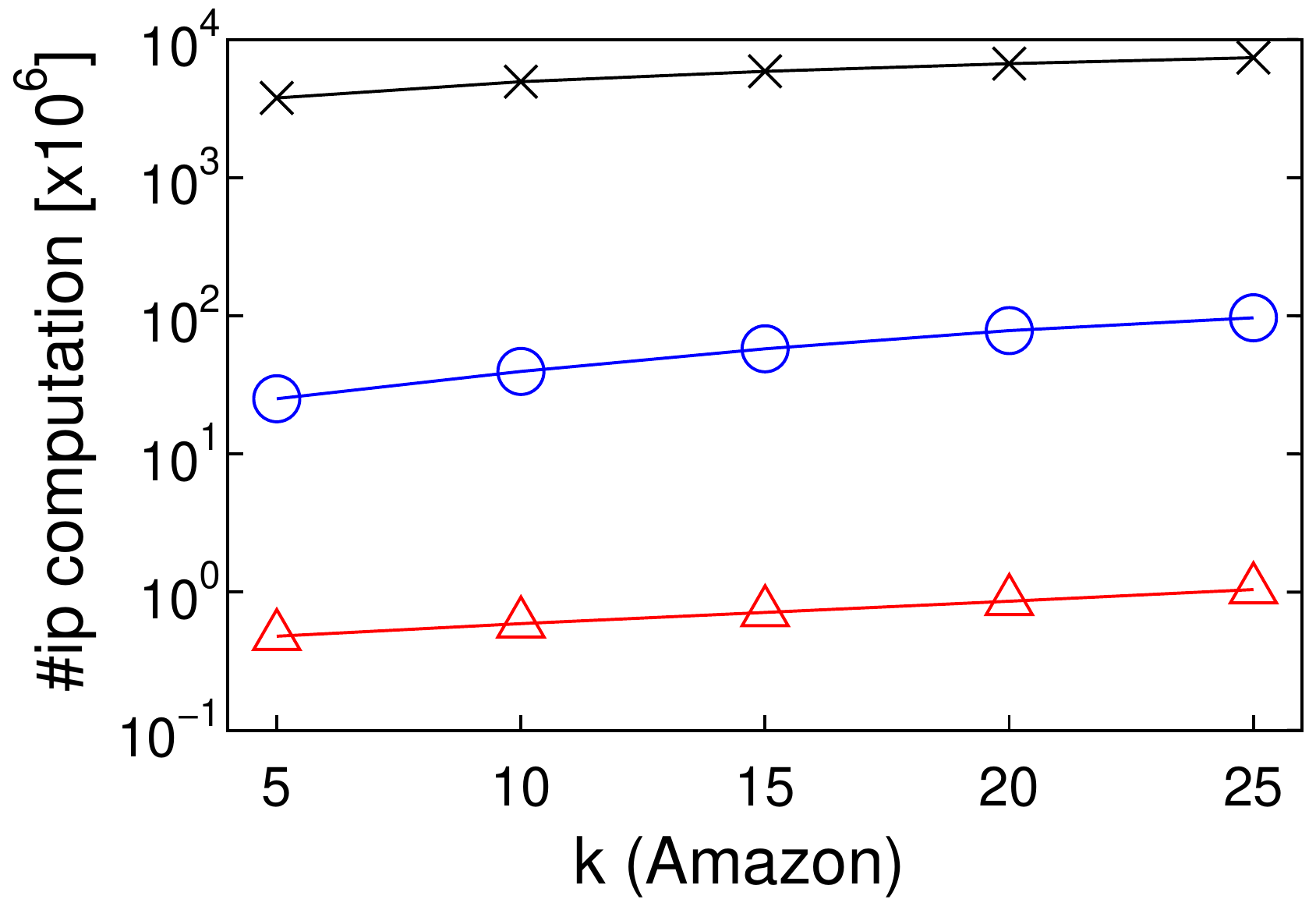}	    \label{figure_amazon-k_ip}}
        \subfigure[Yahoo! (\#ip computations)]{%
		\includegraphics[width=0.24\linewidth]{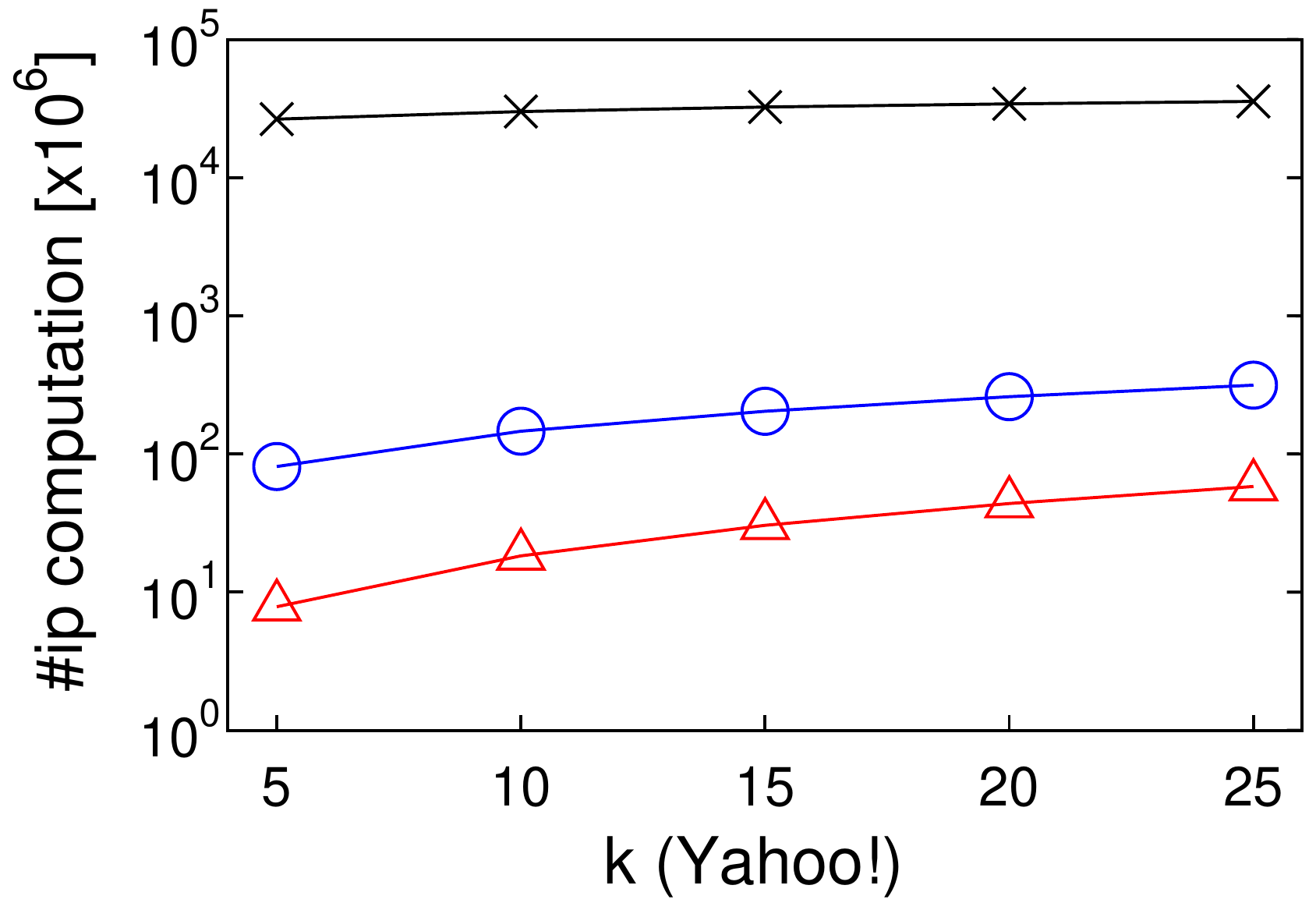}  	    \label{figure_yahoo-k_ip}}
        \caption{Impact of $k$: Running time (top) and \#ip computations (bottom). ``$\times$'' shows LEMP, ``\textcolor{blue}{$\circ$}'' shows FEXIPRO, and ``\textcolor{red}{$\triangle$}'' shows Simpfer.}
        \label{figure_k}
        \Description[Running time of each algorithm increases as $k$ increases]{Running time of each algorithm increases as $k$ increases on MovieLens, Netflix, Amazon, and Yahoo!.}
	\end{center}
\end{figure*}

We investigate how $k$ affects the computational performance of each algorithm by using a single core.
Figure \ref{figure_k} depicts the experimental results.

We first observe that, as $k$ increases, the running time of each algorithm increases, as shown in Figures \ref{figure_movielens-k}--\ref{figure_yahoo-k}.
This is reasonable, because the cost of (decision version of)  $k$-MIPS increases.
As a proof, Figures \ref{figure_movielens-k_ip}--\ref{figure_yahoo-k_ip} show that the number of inner product (ip) computations increases as $k$ increases.
The running time of Simpfer is (sub-)linear to $k$ (the plots are log-scale).
This suggests that $m' = O(k)$.

Second, Simpfer significantly outperforms LEMP and FEXIPRO.
This result is derived from our idea of quickly solving the $k$-MIPS decision problem.
The techniques introduced in Section \ref{section_bounding} can deal with both yes and no answer cases efficiently.
Therefore, our approach functions quite well in practice.

Last, an interesting observation is the performance differences between FEXIPRO and Simpfer.
Let us compare them with regard to running time.
Simpfer \textit{is at least two} orders of magnitude faster than FEXIPRO.
On the other hand, with regard to the number of inner product computations, that of Simpfer is \textit{one} order of magnitude lower than that of FEXIPRO.
This result suggests that the filtering cost of Simpfer is light, whereas that of FEXIPRO is heavy.
Recall that Lemmas \ref{lemma_vector-no}--\ref{lemma_block} need only $O(1)$ time, and Corollaries \ref{corollary_yes}--\ref{corollary_no} need $O(k)$ time in practice.
On the other hand, for each user vector in $Q$, FEXIPRO incurs $\Omega(k)$ time, and its filtering cost is $O(d')$, where $d' < d$.
For high-dimensional vectors, the difference between $O(1)$ and $O(d')$ is large.
From this point of view, we can see the efficiency of Simpfer.

\begin{figure*}[!t]
	\begin{center}
        \subfigure[MovieLens (Running time)]{%
		\includegraphics[width=0.24\linewidth]{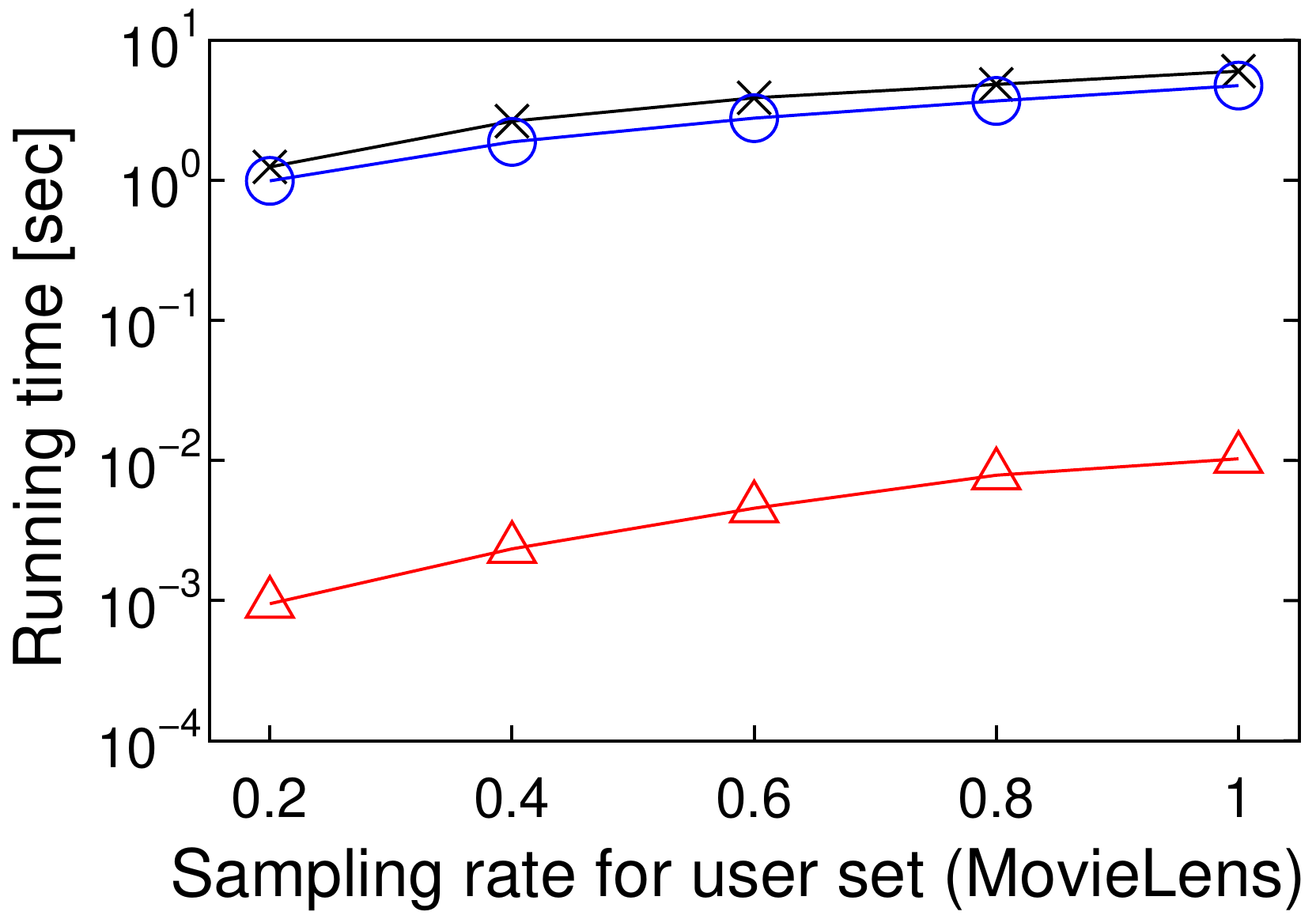}    \label{figure_movielens-cardinality}}
        \subfigure[Netflix (Running time)]{%
		\includegraphics[width=0.24\linewidth]{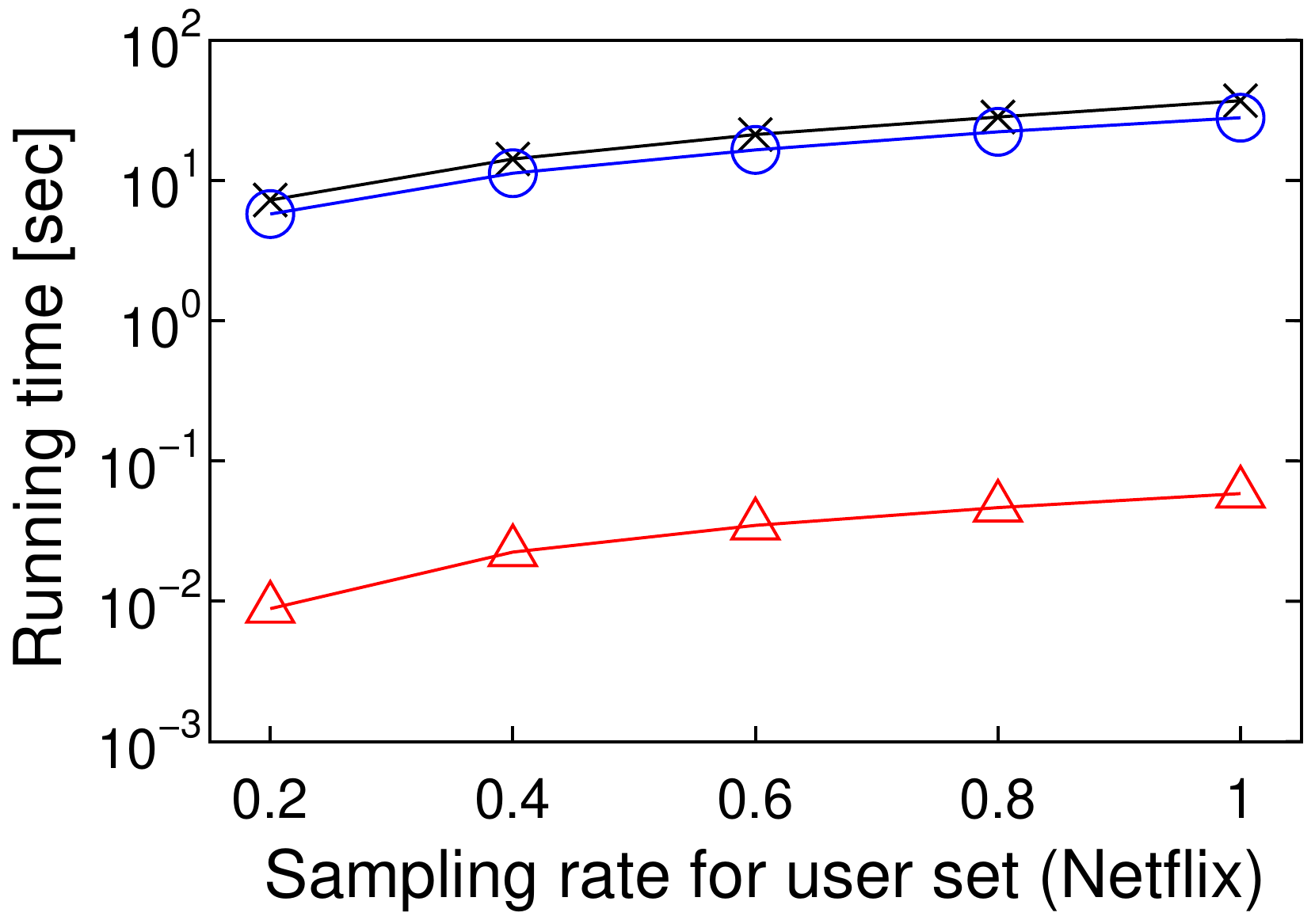}	    \label{figure_netflix-cardinality}}
		\subfigure[Amazon (Running time)]{%
		\includegraphics[width=0.24\linewidth]{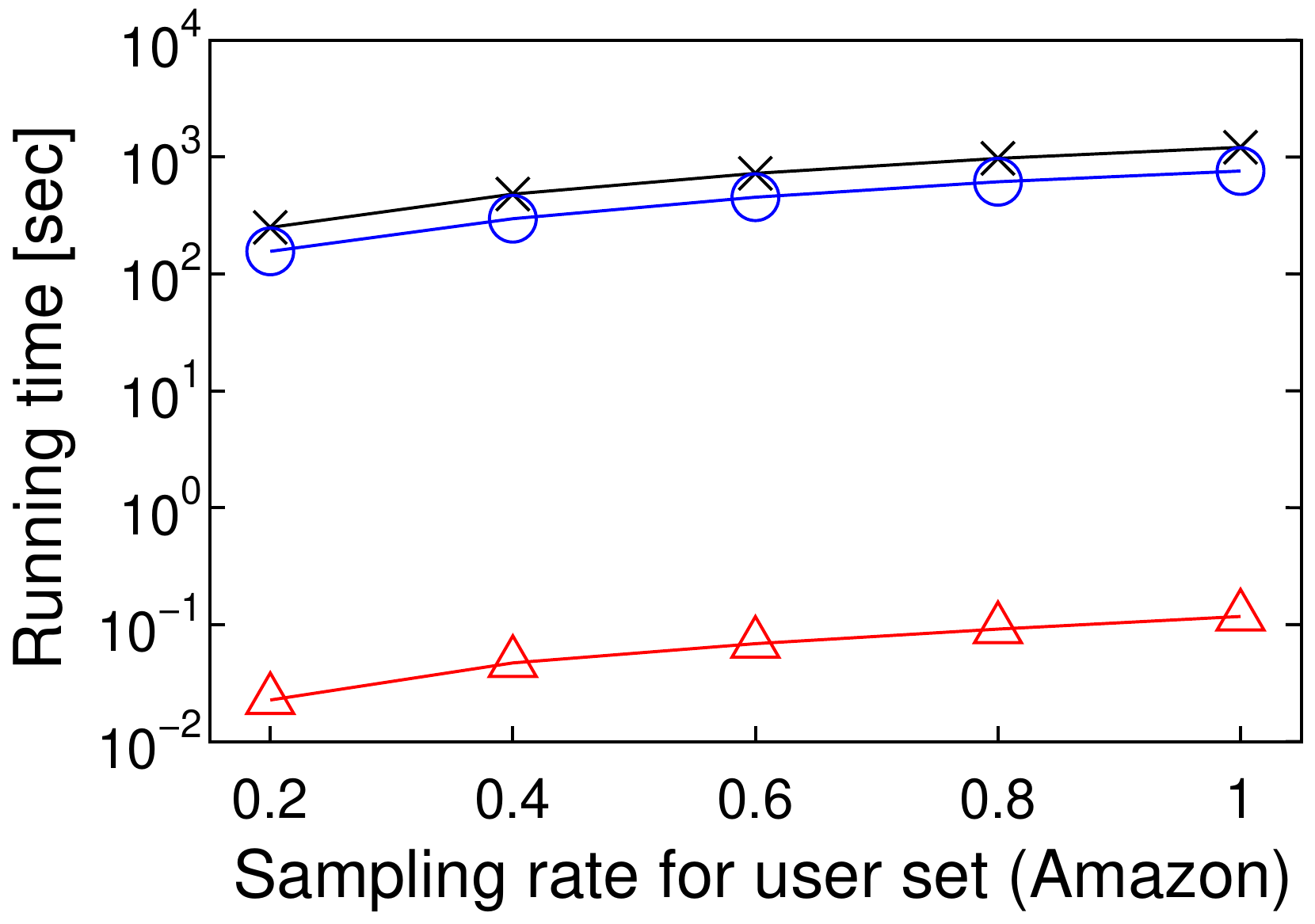}	    \label{figure_amazon-cardinality}}
        \subfigure[Yahoo! (Running time)]{%
		\includegraphics[width=0.24\linewidth]{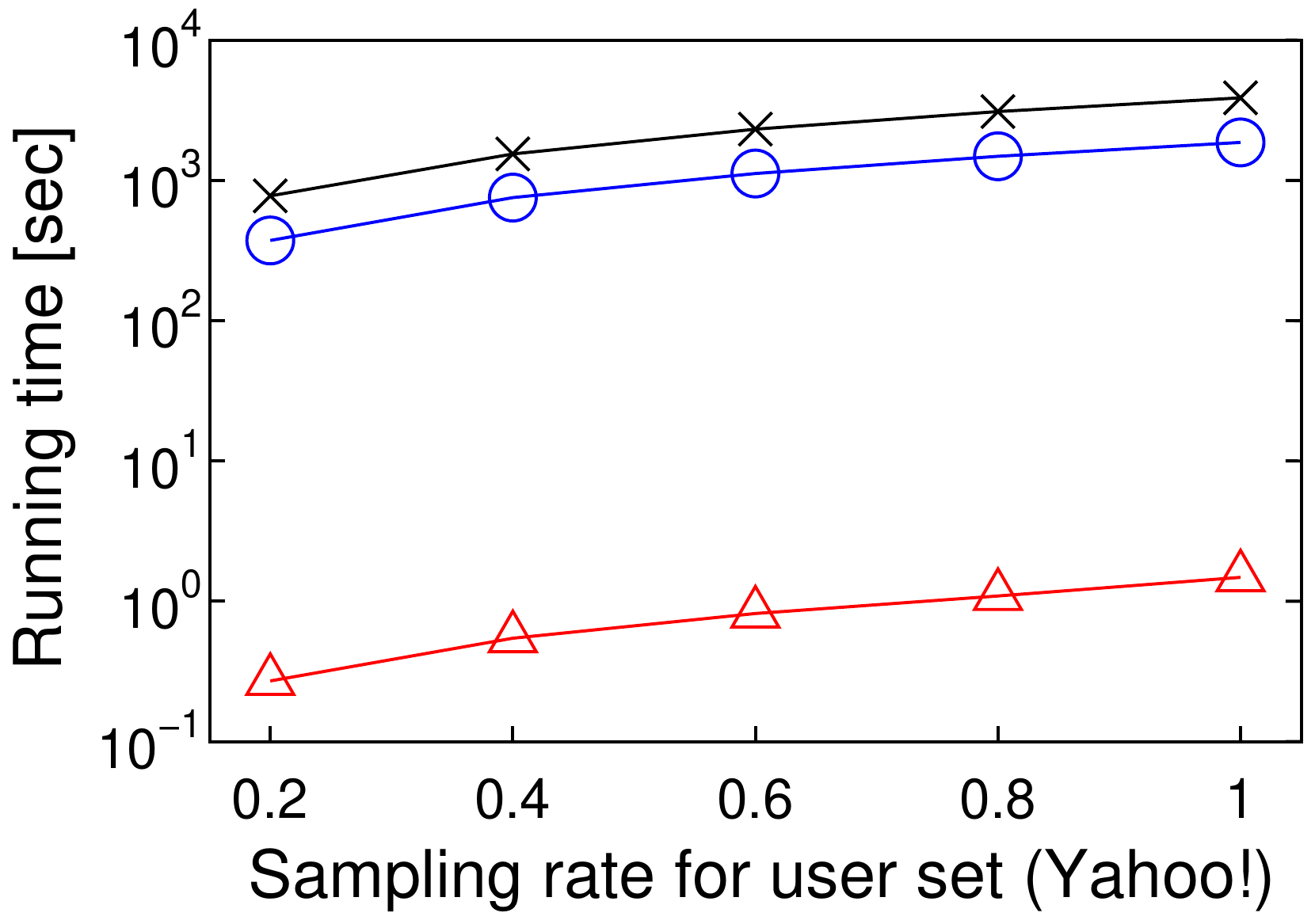}  	    \label{figure_yahoo-cardinality}}
		\subfigure[MovieLens (\#ip computations)]{%
		\includegraphics[width=0.24\linewidth]{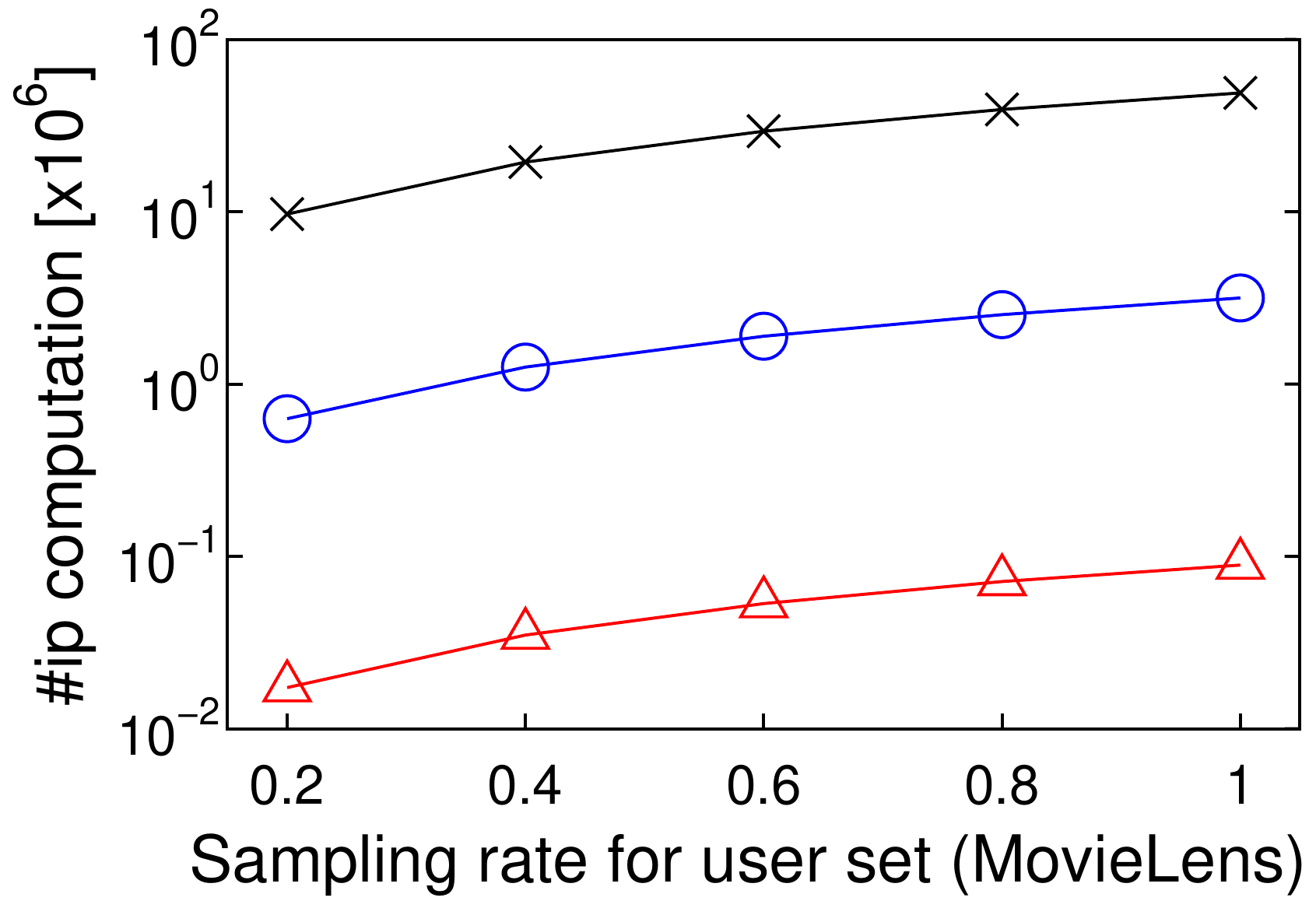} \label{figure_movielens-cardinality_ip}}
        \subfigure[Netflix (\#ip computations)]{%
		\includegraphics[width=0.24\linewidth]{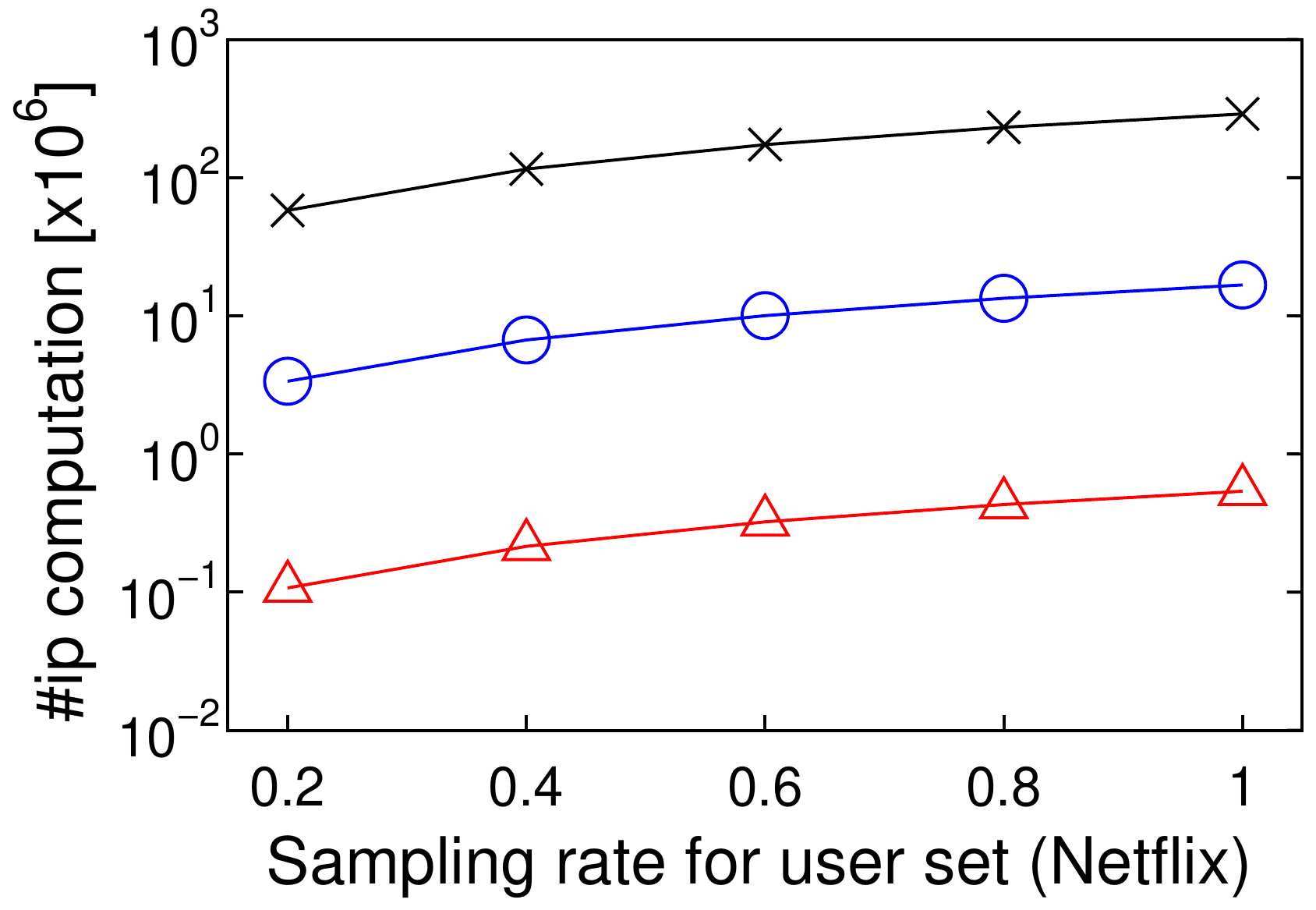}	\label{figure_netflix-cardinality_ip}}
		\subfigure[Amazon (\#ip computations)]{%
		\includegraphics[width=0.24\linewidth]{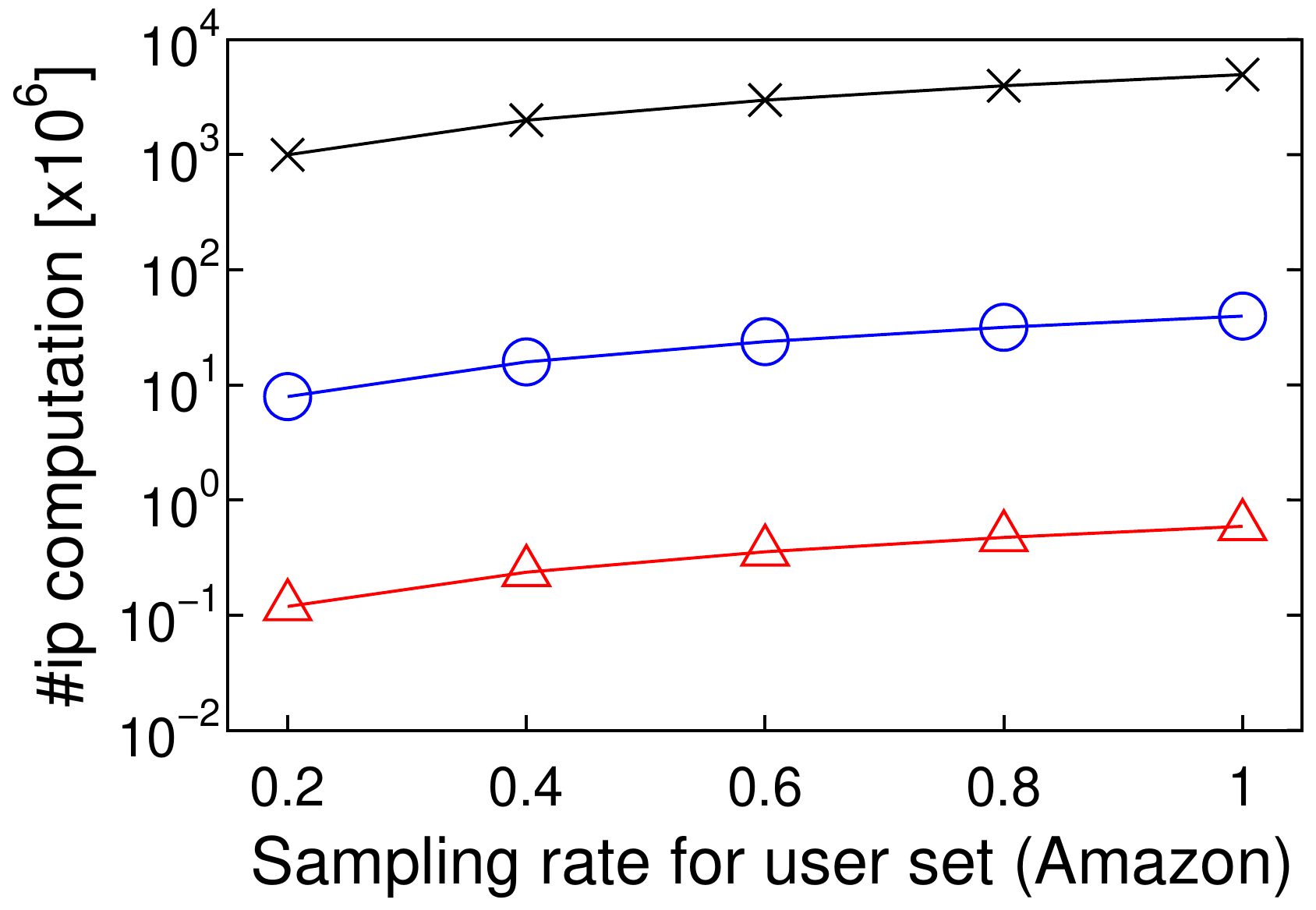}	\label{figure_amazon-cardinality_ip}}
        \subfigure[Yahoo! (\#ip computations)]{%
		\includegraphics[width=0.24\linewidth]{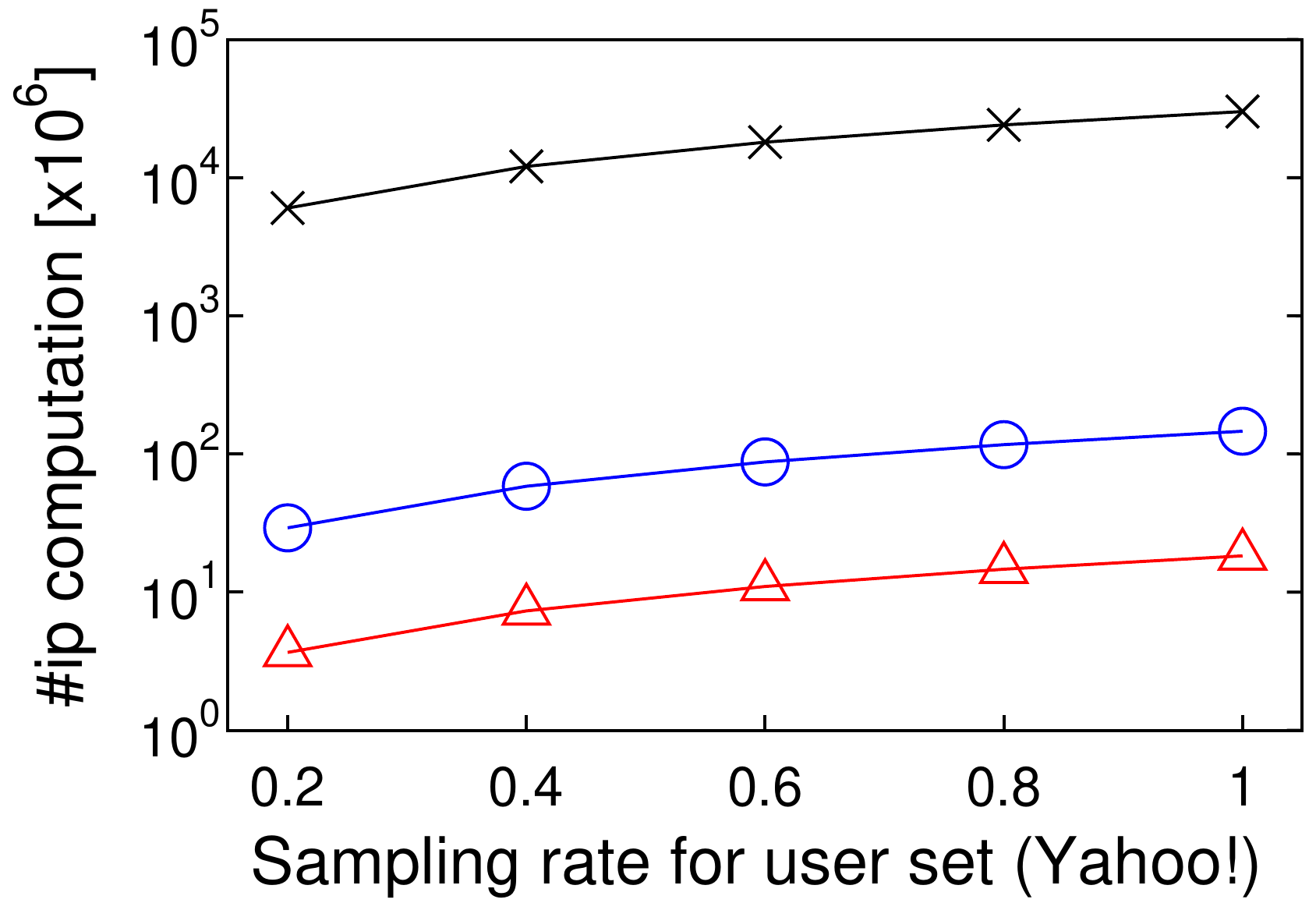}  	\label{figure_yahoo-cardinality_ip}}
        \caption{Impact of $|\mathbf{Q}|$: Running time (top) and \#ip computations (bottom). ``$\times$'' shows LEMP, ``\textcolor{blue}{$\circ$}'' shows FEXIPRO, and ``\textcolor{red}{$\triangle$}'' shows Simpfer.}
        \Description[Running time of each algorithm increases as the cardinality of user vector set increases]{Running time of each algorithm increases as the cardinality of user vector set increases on MovieLens, Netflix, Amazon, and Yahoo!.}
        \label{figure_cardinality}
	\end{center}
\end{figure*}

\subsection{Result 3: Impact of Cardinality of $\mathbf{Q}$}
We next study the scalability to $n = |\mathbf{Q}|$ by using a single core.
To this end, we randomly sampled $s \times n$ user vectors in $\mathbf{Q}$, and this sampling rate $s$ has $s \in [0.2,1.0]$.
We set $k = 10$.
Figure \ref{figure_cardinality} shows the experimental result.

In a nutshell, we have a similar result to that in Figure \ref{figure_k}.
As $n$ increases, the running time of Simpfer linearly increases.
This result is consistent with Theorem \ref{theorem_time}.
Notice that the tendency of the running time of Simpfer follows that of the number of inner product computations.
This phenomenon is also supported by Theorem \ref{theorem_time}, because the main bottleneck of Simpfer is \textsc{Linear-scan}$(\cdot)$.

\begin{figure*}[!t]
	\begin{center}
        \subfigure[MovieLens (Running time)]{%
		\includegraphics[width=0.24\linewidth]{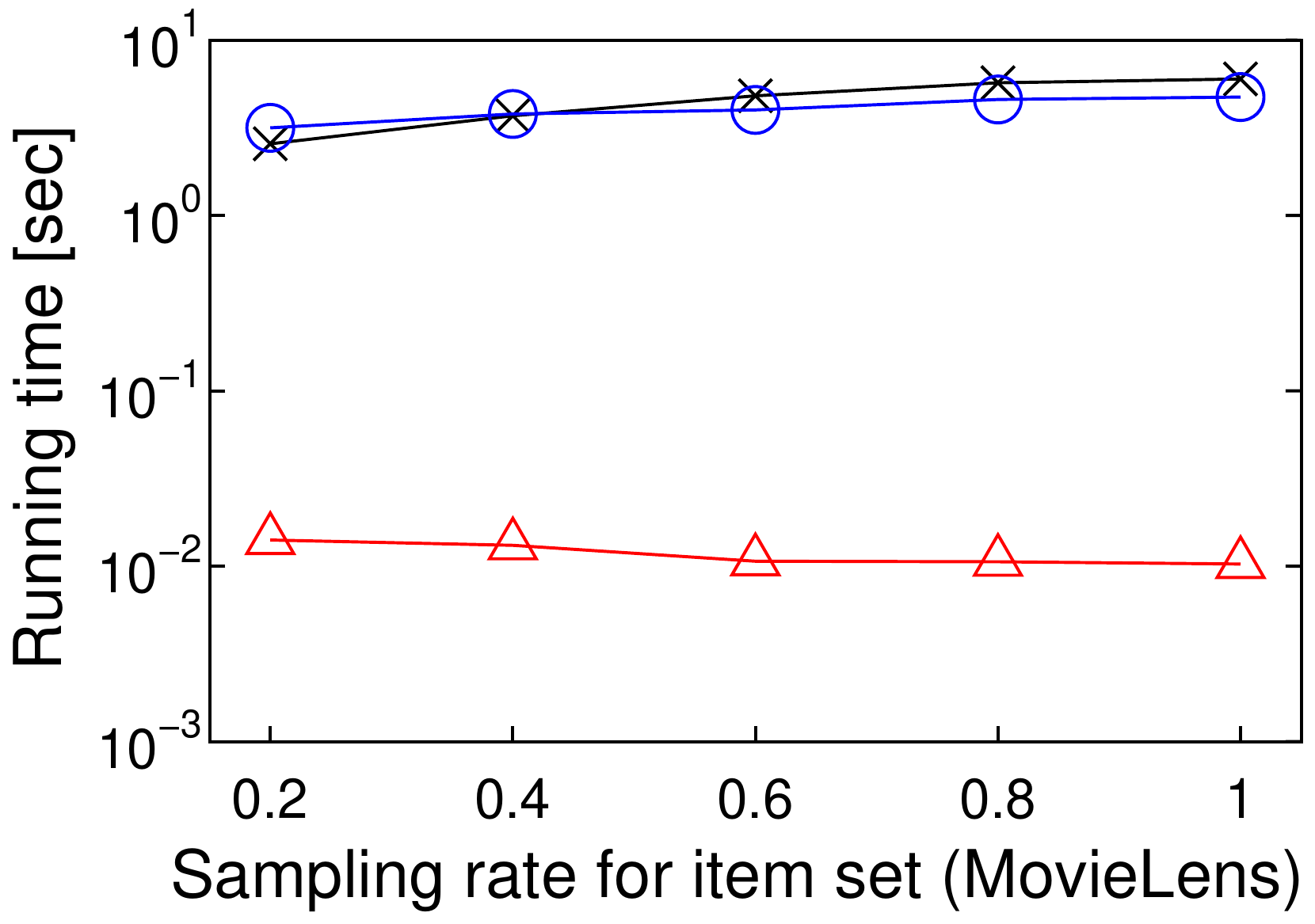}    \label{figure_movielens-cardinality_}}
        \subfigure[Netflix (Running time)]{%
		\includegraphics[width=0.24\linewidth]{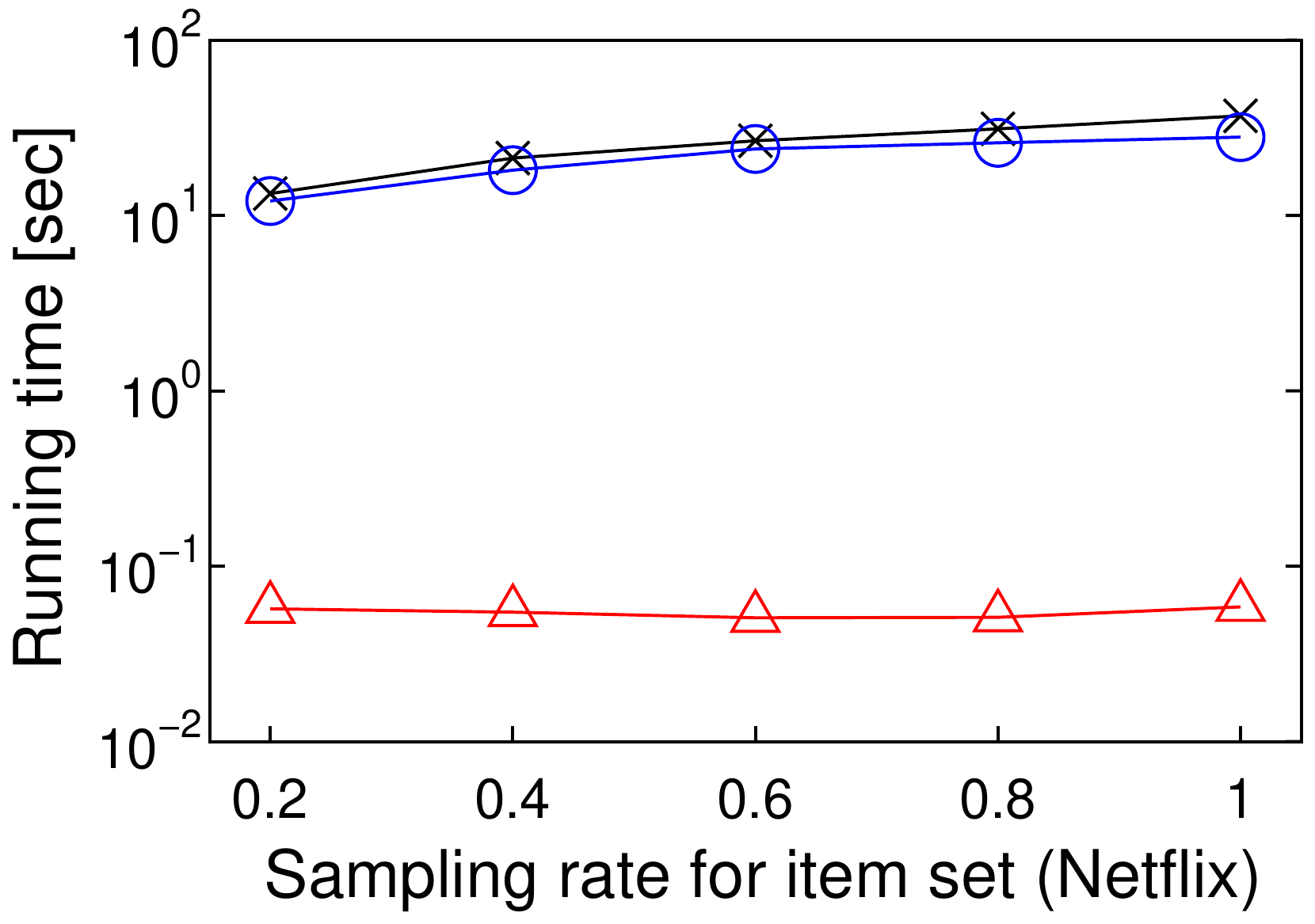}	    \label{figure_netflix-cardinality_}}
		\subfigure[Amazon (Running time)]{%
		\includegraphics[width=0.24\linewidth]{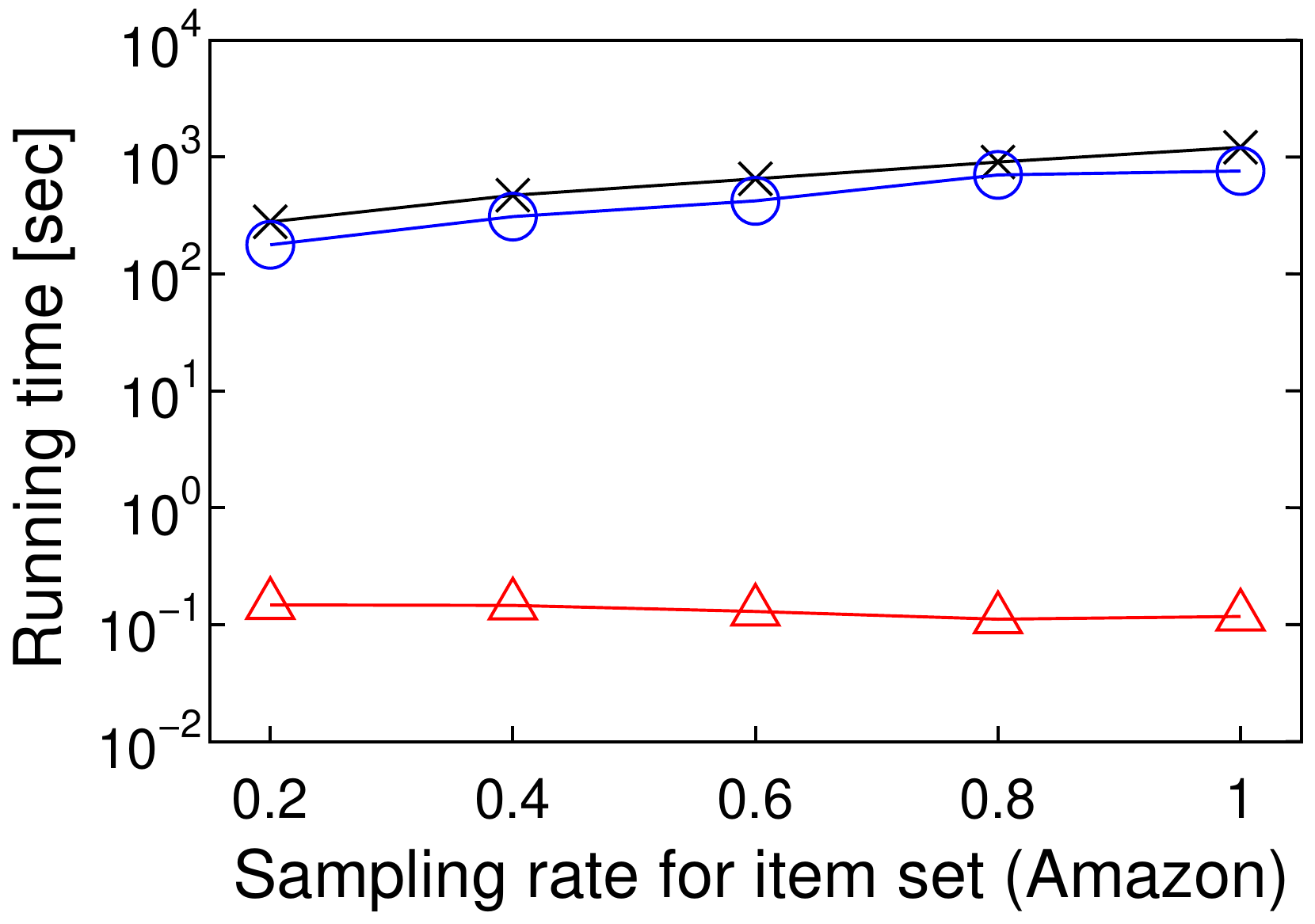}	    \label{figure_amazon-cardinality_}}
        \subfigure[Yahoo! (Running time)]{%
		\includegraphics[width=0.24\linewidth]{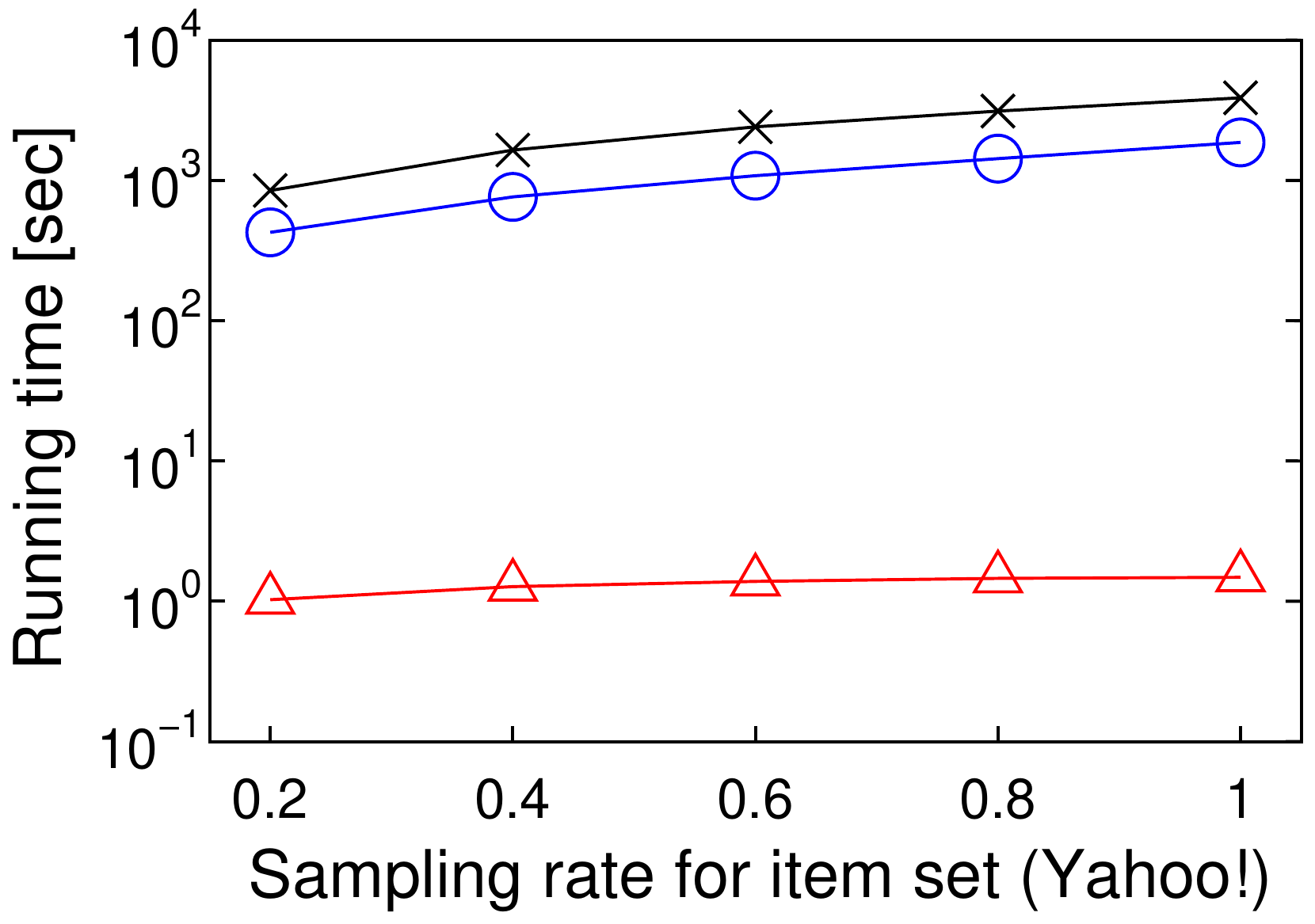}  	    \label{figure_yahoo-cardinality_}}
		\subfigure[MovieLens (\#ip computations)]{%
		\includegraphics[width=0.24\linewidth]{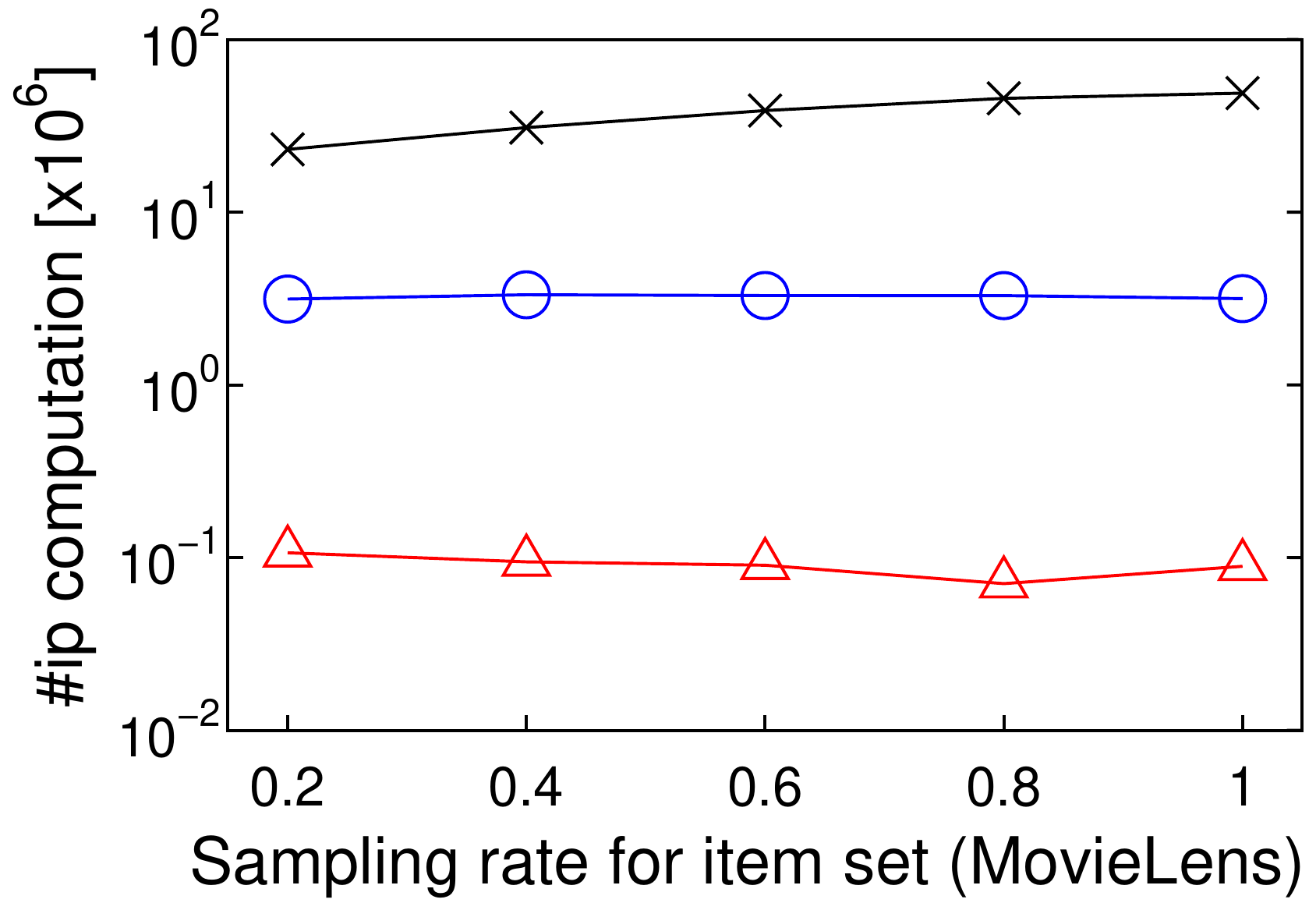} \label{figure_movielens-cardinality_ip_}}
        \subfigure[Netflix (\#ip computations)]{%
		\includegraphics[width=0.24\linewidth]{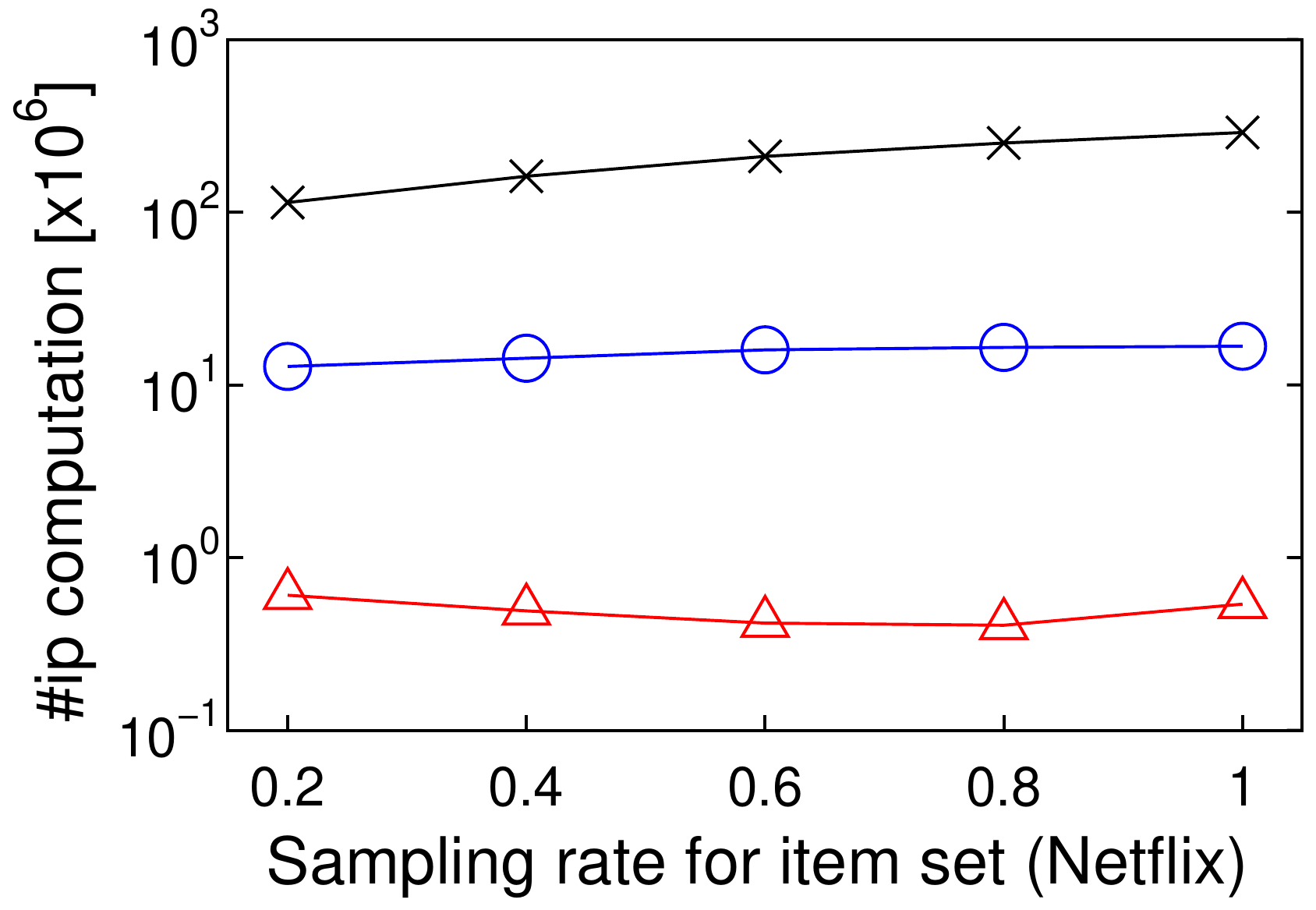}	\label{figure_netflix-cardinality_ip_}}
		\subfigure[Amazon (\#ip computations)]{%
		\includegraphics[width=0.24\linewidth]{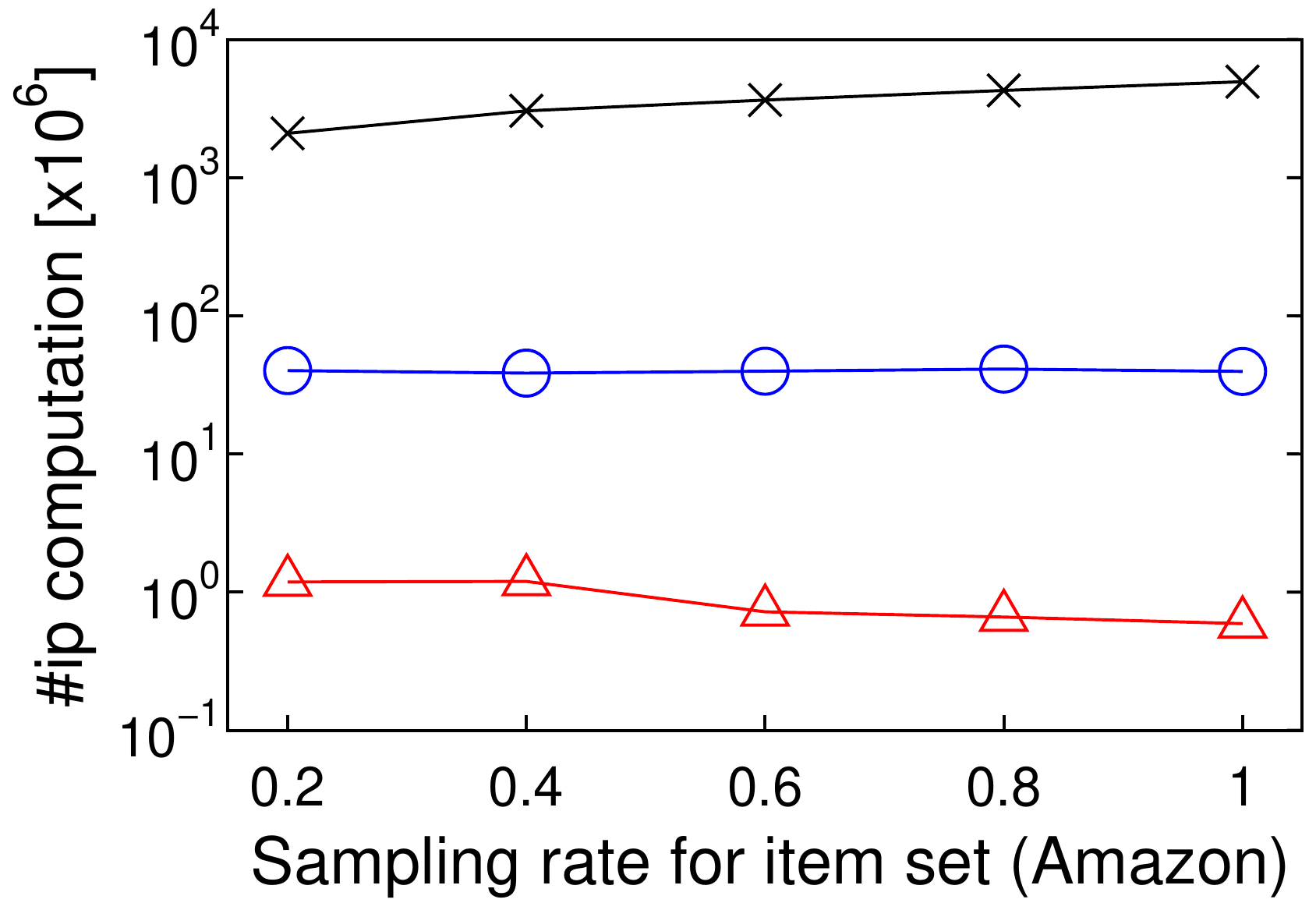}	\label{figure_amazon-cardinality_ip_}}
        \subfigure[Yahoo! (\#ip computations)]{%
		\includegraphics[width=0.24\linewidth]{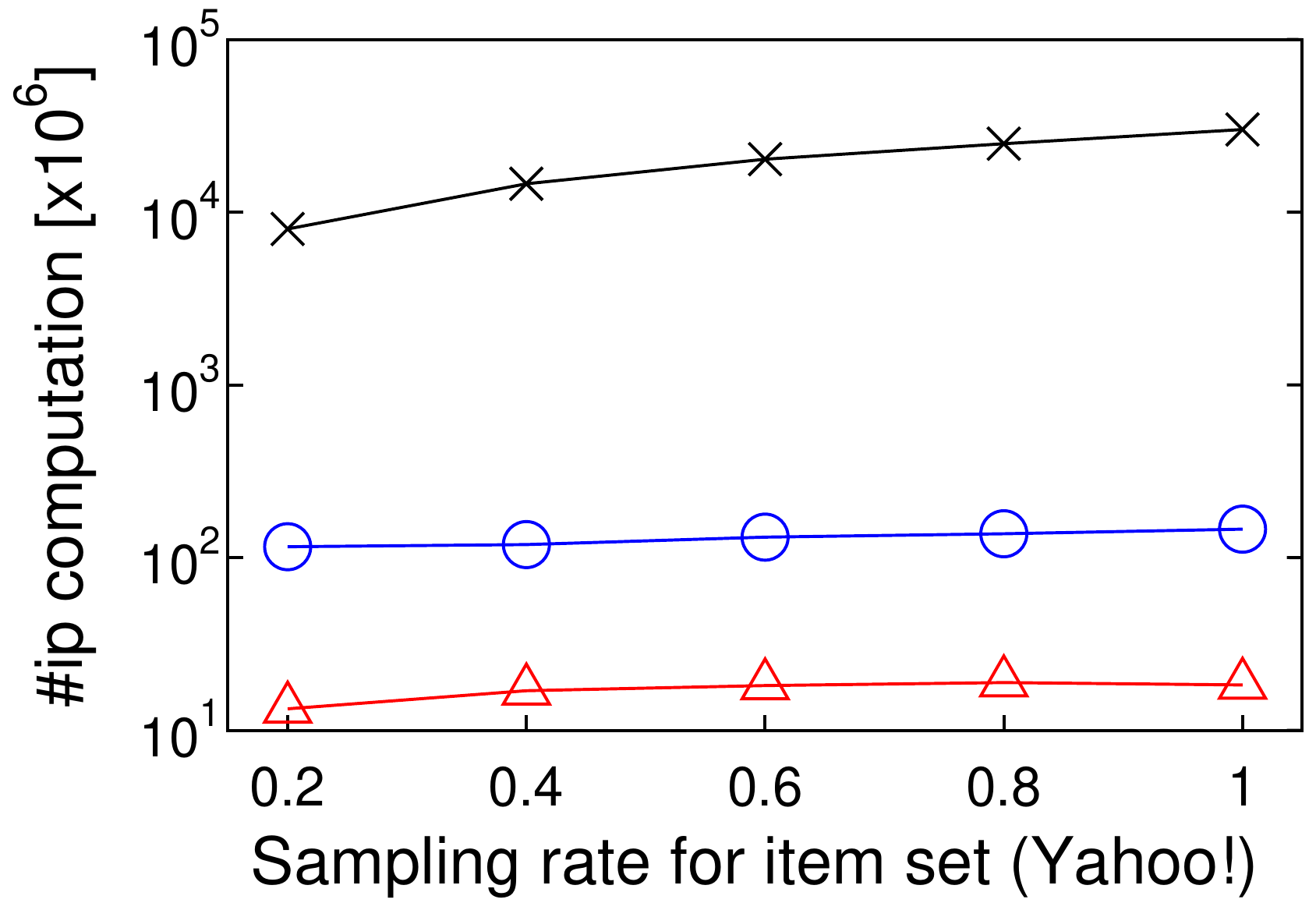}  	\label{figure_yahoo-cardinality_ip_}}
        \caption{Impact of $|\mathbf{P}|$: Running time (top) and \#ip computations (bottom). ``$\times$'' shows LEMP, ``\textcolor{blue}{$\circ$}'' shows FEXIPRO, and ``\textcolor{red}{$\triangle$}'' shows Simpfer.}
        \label{figure_cardinality_}
        \Description[Running time of Simpfer does not increases as the cardinality of item vector set increases]{Running time of Simpfer does not increases as the cardinality of item vector set increases on MovieLens, Netflix, Amazon, and Yahoo!.}
	\end{center}
\end{figure*}

\subsection{Result 4: Impact of Cardinality of $\mathbf{P}$}
The scalability to $m = |\mathbf{P}|$ by using a single core is also investigated.
We randomly sampled $s \times m$ user vectors in $\mathbf{P}$, as with the previous section.
Figure \ref{figure_cardinality_} shows the experimental result where $k = 10$.
Interestingly, we see that the result is different from that in Figure \ref{figure_cardinality}.
The running time of Simpfer is almost stable for different $m$.
In this experiment, $n$ and $k$ were fixed, and recall that $m' = O(k)$.
From this observation, the stable performance is theoretically obtained.
This scalability of Simpfer is an advantage over the other algorithms, since their running time increases as $m$ increases.

\subsection{Result 5: Impact of Number of CPU Cores}    \label{section_experiment-multicore}
We study the gain of multicore processing of Simpfer by setting $k = 10$.
We depict the speedup ratios compared with the single-core case in Table \ref{table_core}.

\begin{table}[!h]
    \centering
	\caption{Speedup ratios of Simpfer}
	\label{table_core}
	\Description[Multi-threading improves the efficiency of Simpfer]{Multi-threading improves the efficiency of Simpfer on MovieLens, Netflix, Amazon, and Yahoo!.}
	\begin{tabular}{c|c|c|c|c} \hline
        \#cores     &   MovieLens   & Netflix   & Amazon    & Yahoo!    \\ \hline \hline
        4           &   3.23        & 2.99      & 3.41      & 2.22      \\ \hline
        8           &   4.80        & 4.27      & 6.61      & 2.83      \\ \hline
        12          &   5.84        & 5.40      & 7.76      & 2.88      \\ \hline
	\end{tabular}
\end{table}

We see that Simpfer receives the benefit of multicore processing, and its running time decreases as the number of available cores increases.
We here explain why Simpfer cannot obtain speedup ratio $c$, where $c$ is the number of available cores.
Each core deals with different blocks, and the processing cost of a given block $\mathbf{B}$ is different from those of the others.
This is because it is unknown whether $\mathbf{B}$ can be pruned by Lemma \ref{lemma_block}.
Even if we magically know the cost, it is NP-hard to assign blocks so that each core has the same processing cost \cite{amagata2019identifying, korf2009multi}.
Therefore, perfect load-balancing is impossible in practice.
The Yahoo! case in particular represents this phenomenon.
Because many user vectors in Yahoo! have large norms, blocks often cannot be filtered by Lemma \ref{lemma_block}.
This can be seen from the observation in Figure \ref{figure_yahoo-cardinality_ip}: the number of inner product computations on Yahoo! is larger than those on the other datasets.
The costs of Corollaries \ref{corollary_yes}--\ref{corollary_no} are data-dependent (i.e., they are not pre-known), rendering a fact that Yahoo! is a hard case for obtaining a high speedup ratio.

\begin{table}[!h]
    \centering
	\caption{Pre-processing time of Simpfer [sec]}
	\label{table_pre-processing-time}
	\Description[Pre-processing time of Simpfer is reasonable]{Pre-processing time of Simpfer is reasonable on MovieLens, Netflix, Amazon, and Yahoo!.}
	\begin{tabular}{c|c|c|c} \hline
        MovieLens   & Netflix   & Amazon    & Yahoo!    \\ \hline \hline
        1.02        & 4.08      & 15.10     & 15.55     \\ \hline
	\end{tabular}
\end{table}

\subsection{Result 6: Pre-processing Time}  \label{section_experiment-offline}
Last, we report the pre-processing time of Simpfer.
Table \ref{table_pre-processing-time} shows the results.
As Theorem \ref{theorem_pre-time} demonstrates, the pre-processing time increases as $n$ increases.
We see that the pre-processing time is reasonable and much faster than the online (running) time of the baselines.
For example, the running time of FEXIPRO on Amazon with $k = 25$ is 1206 [sec].
When $k = 25$ (i.e., $k = k_{max}$), the total time of pre-processing and online processing of Simpfer is $15.10 + 0.16 = 15.26$ [sec].
Therefore, even if $k > k_{max}$ is specified, re-building blocks then processing the query by Simpfer is much faster.

\section{Conclusion}    \label{section_conclusion}
This paper introduced a new problem, reverse maximum inner product search (reverse MIPS).
The reverse MIPS problem supports many applications, such as recommendation, advertisement, and market analysis.
Because even state-of-the-art algorithms for MIPS cannot solve the reverse MIPS problem efficiently, we proposed Simpfer as an exact and efficient solution.
Simpfer exploits several techniques to efficiently answer the decision version of the MIPS problem.
Our theoretical analysis has demonstrated that Simpfer is always better than a solution that employs a state-of-the-art algorithm of MIPS.
Besides, our experimental results on four real datasets show that Simpfer is at least two orders of magnitude faster than the MIPS-based solutions.

\begin{acks}
This research is partially supported by JST PRESTO Grant Number JPMJPR1931, JSPS Grant-in-Aid for Scientific Research (A) Grant Number 18H04095, and JST CREST Grant Number JPMJCR21F2.
\end{acks}

\bibliographystyle{ACM-Reference-Format}
\bibliography{sigproc}

\end{document}